\documentclass[twocolumn]{aastex7}

\usepackage{graphicx}

\usepackage{amsmath}
\usepackage{url}
\usepackage{subfigure}
\usepackage{xspace}

\shorttitle{The Disrupted Cluster OCSN-49}
\shortauthors{Miller et al.}
\newcommand{\gaia}{{Gaia}\xspace}

\begin{document}
\title{Evidence for a Catastrophically Disrupted Open Cluster}
\correspondingauthor{Alexis Miller}
\email{alexis.miller@spartans.ut.edu}

\author[0009-0008-8119-1115]{Alexis N. Miller}
\affiliation{Department of Physics and Astronomy, University of Tampa, Tampa, FL 33606, USA}
\email{alexis.miller@spartans.ut.edu}

\author[0009-0009-6352-4964]{Kyle R. Tregoning}
\affiliation{Department of Physics, University of Florida, 2001 Museum Rd, Gainesville, FL 32611, USA}
\email{kyle.tregoning@ufl.edu}

\author[0000-0001-5261-3923]{Jeff J. Andrews}
\affiliation{Department of Physics, University of Florida, 2001 Museum Rd, Gainesville, FL 32611, USA}
\affiliation{Institute for Fundamental Theory, 2001 Museum Rd, Gainesville, FL 32611, USA}
\email{jeffrey.andrews@ufl.edu}

\author[0000-0001-7203-8014]{Simon C. Schuler}
\affiliation{Department of Physics and Astronomy, University of Tampa, Tampa, FL 33606, USA}
\email{sschuler@ut.edu}

\author[0000-0002-2792-134X]{Jason L.~Curtis}
\affiliation{Department of Astronomy, Columbia University, 550 West 120th Street, New York, NY 10027, USA}
\email{jasoncurtis.astro@gmail.com}

\author[0000-0001-7077-3664]{Marcel A.~Ag\"{u}eros}
\affiliation{Department of Astronomy, Columbia University, 550 West 120th Street, New York, NY 10027, USA}
\affiliation{Laboratoire d’astrophysiquede Bordeaux, Univ. Bordeaux, CNRS, B18N, Allée Geoffroy Saint-Hilaire, 33615 Pessac, France}
\email{marcel@astro.columbia.edu}

\author[0000-0002-1617-8917]{Phillip A.~Cargile}
\affiliation{Center for Astrophysics | Harvard \& Smithsonian, 60 Garden Street, Cambridge, MA 02138, USA}
\email{phillip.cargile@cfa.harvard.edu}

\author[0000-0003-2481-4546]{Julio Chanam{\'e}}
\affiliation{Instituto de Astrofísica, Pontificia Universidad Católicade Chile, Av. Vicuña Mackenna 4860, 782-0436 Macul, Santiago, Chile}
\email{jchaname@astro.puc.cl}

\begin{abstract}
Of the many discoveries uncovered by the \gaia astrometric mission, some of the most exciting are related to nearby dispersed stellar structures. We analyze one such structure in the Milky Way disk, OCSN-49, a coeval stellar stream with 257 identified members spanning approximately 30$^{\circ}$ across the sky. We obtained high-resolution spectroscopic data for four members that span the stream's extent, finding these four stars to have solar metallicities and remarkably homogeneous chemistry. Through a combination of isochrone fitting, lithium abundance analysis, and gyrochronology, we find a consistent stellar age of 400--600 Myr. Integrating stellar orbits backwards reveals that OCSN-49 converged to a single point at a much younger age. By integrating unbound model stars forward and comparing them to the current phase space distribution of OCSN-49, we derive a dynamical age of 83 $\pm$ 1 Myr, inconsistent with the age of the stellar population. The discrepancy between the kinematic and stellar age indicators is naturally explained by a disruptive event that unbound OCSN-49 roughly 500 Myr into its lifetime. Based on rate estimates, disruption due to a passing giant molecular cloud (GMC) is the most likely culprit. Assuming a single encounter, we find that a nearly head-on collision with a fairly massive GMC ($\sim 10^5$ M$_{\odot}$) was necessary to unbind the cluster, although encounters with multiple GMCs may be responsible. To our knowledge, OCSN-49 serves as the first known remnant of a catastrophically disrupted open cluster and therefore serves as a benchmark for further investigating cluster disruption in the Milky Way.
\end{abstract}

\section{Introduction} \label{sec:intro}
The European Space Agency (ESA) \gaia mission catalog is revolutionizing our knowledge of stellar populations throughout the Milky Way \citep{GaiaShedsLight}. The third data release includes 1.5 billion sources with precise five-dimensional (5D) astrometry, roughly 33.8 million of which include radial velocity measurements \citep{GaiaDR3}. By probing these data, a multitude of previously unknown stellar structures have been uncovered including Meingast 1, the most massive stellar stream in the solar neighborhood \citep{Meingast1}, 34 open clusters around the Galactic bulge \citep{Ferreira}, and the large and luminous cluster Gaia 1, positioned behind the bright star Sirius \citep{Koposov}.

These groups of coeval stars provide a wealth of information about star formation and stellar evolution, as well as the history of the Milky Way itself. It is commonly assumed that most stars are not born independently but are formed in clusters embedded within molecular clouds \citep{Lada}, and only an estimated 4\% of those clusters survive to ages greater than 100 Myr in the solar neighborhood \citep{Goodwin_2006, Bastian_2007}. Upon cluster disruption, its component stars travel away from their birth environments and disperse into the Milky Way field \citep{Coronado}. The evolution of stars from birth to their place in the field is not fully understood, but is an active area of study \citep{Kamdar2019a, Helmi}. 

Recently, the application of machine learning algorithms to \gaia astrometric data has offered the tantalizing possibility to observationally study the nature of star cluster disruption in the Milky Way disk. Multiple groups have identified stellar structures within the Milky Way disk by using these machine learning methods to find groups of comoving stars \citep{Kounkel_2019, Qin_2023, HuntCatalog}. While some of these stellar associations may be the remnants of disrupted star clusters, others may be the result of temporary overdensities due to orbital instabilities. To confirm the common origin of a stellar group, multiple methods can be used: isochrone fitting \citep[e.g.,][]{Jorgensen_2005, M67}, gyrochronology \citep{Skumanich_1972, Kawaler_1989, Barnes_2003, Curtis_2020}, and lithium dating can all demonstrate a common age \citep[see][for a review]{ageofstars}. Consistency in the detailed abundance pattern demonstrates the stars' common chemistry \citep{Freeman, Ting}, and precisely measured positions and velocities can be used to show stars share a dynamical origin \citep[e.g.,][]{Wright_2018, Galli_2023}. These conclusions are most robust when multiple methods are combined \citep[e.g.,][]{Schuler}.

Low density, comoving stellar associations in the solar neighborhood, in particular, offer key insights into the transition from clustered environments at birth to the Milky Way field population. Theia 456 (COIN-Gaia-13) \citep{Cantat-Gaudin_2019, Kounkel_2019}, a diffuse stream of stars in the Milky Way disk, provides an example of this transitory phase. \cite{Andrews_2022} robustly showed the structure to be coeval through a combination of chemistry, gyrochronology, and kinematics. \cite{Kyle} derived a precise age estimate of $245 \pm 3$~Myr, as well as birth conditions (scale and velocity dispersion) by integrating model stars forward under a model Milky Way gravitational potential and comparing them to the observed phase space of the structure today. They showed that the present day phase space can be reproduced by a low-density ball of stars evolving under the Galactic tidal field alone, and that the structure will continue to disperse into the field under this tide.

\begin{figure*}
\begin{center}
    \includegraphics[width=1.0\textwidth]{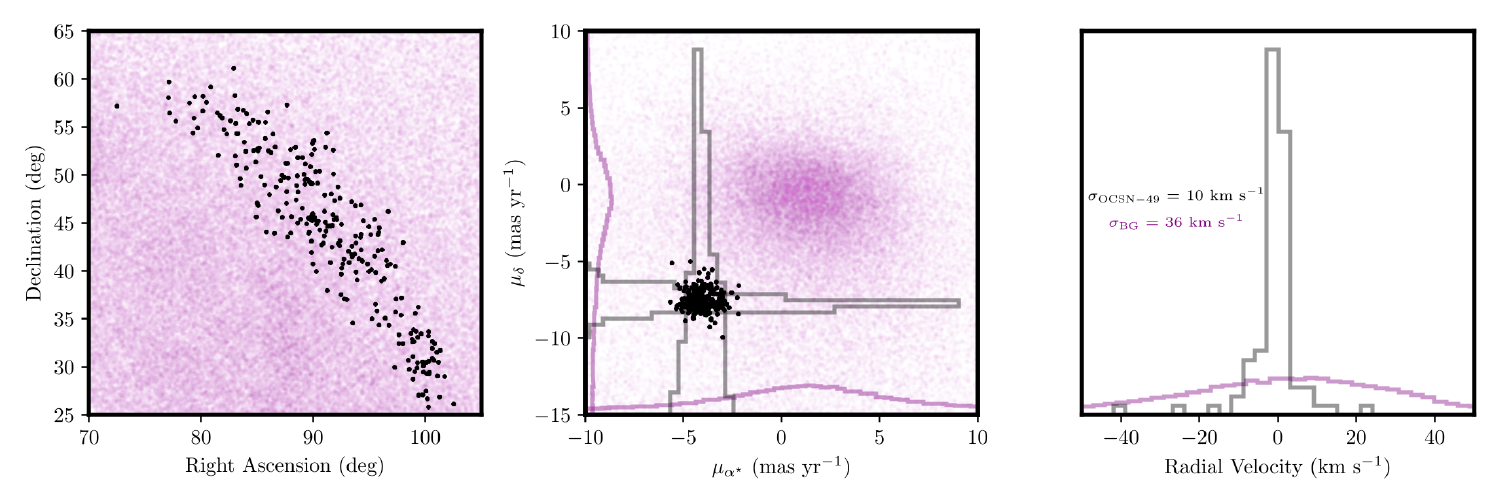}
    \caption{\textit{Left}: \gaia DR3 positions for stars in our dynamical sample (black) and background stars (purple), selected with right ascensions between 70--105$^{\circ}$, declinations between 25--65$^{\circ}$, and parallaxes between 3.85--5.12 mas. \textit{Middle}: Proper motion space of the same stars. Density histograms for OCSN-49 and the background stars are plotted for $\mu_{\alpha^{\star}}$ and $\mu_{\delta}$, all with binwidths of $0.4$ mas~yr$^{-1}$. \textit{Right}: Density histograms for stars with \gaia radial velocities for both OCSN-49 and background stars, each with a binwidth of $3$ km~s$^{-1}$. The standard deviation in radial velocities for both populations is also displayed. OCSN-49 stands out clearly as an overdensity in both proper motion and radial velocity space. }\label{background}
\end{center}
\end{figure*}

\begin{figure}
   \includegraphics[width=1.0\columnwidth]{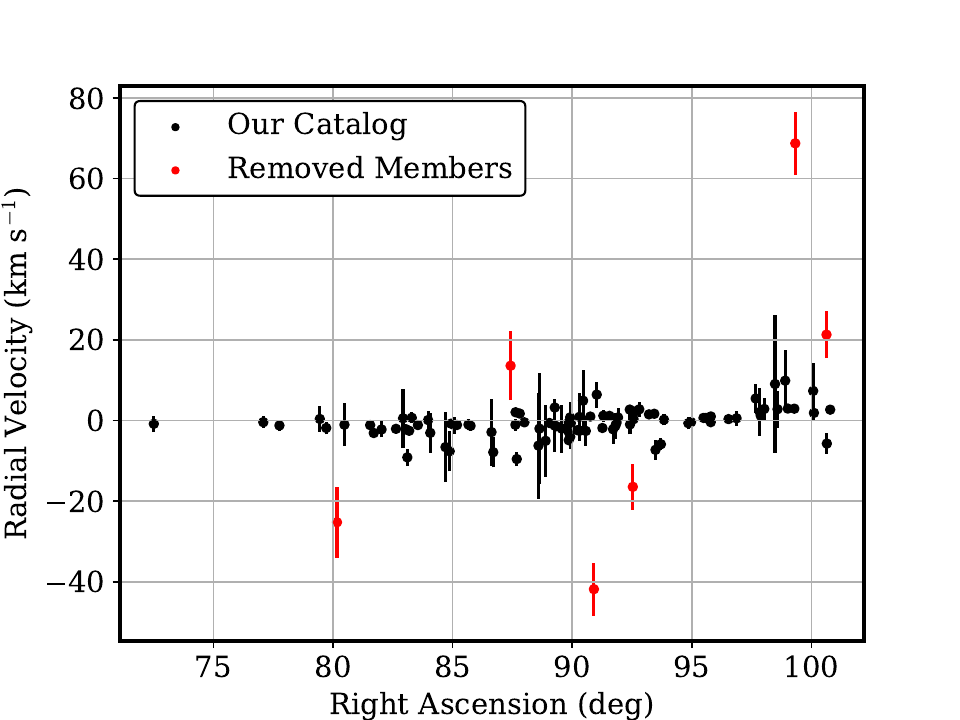}
   \caption{Right ascension versus \gaia radial velocity of OCSN-49 members identified by the \cite{HuntCatalog} catalog. Red circle markers denote OCSN-49 members with radial velocities deviating by more than 10 km s$^{-1}$ from the median that do not appear to be astrometric binaries. We cull these members from our sample. The black circles denote the stars with measured radial velocities included in our final sample. We note the two stars that we culled via proper motions do not have \gaia radial velocities, so they are not included in this plot.}\label{sample}
\end{figure}

\begin{figure}
\begin{center}

   \includegraphics[width=1.0\columnwidth]{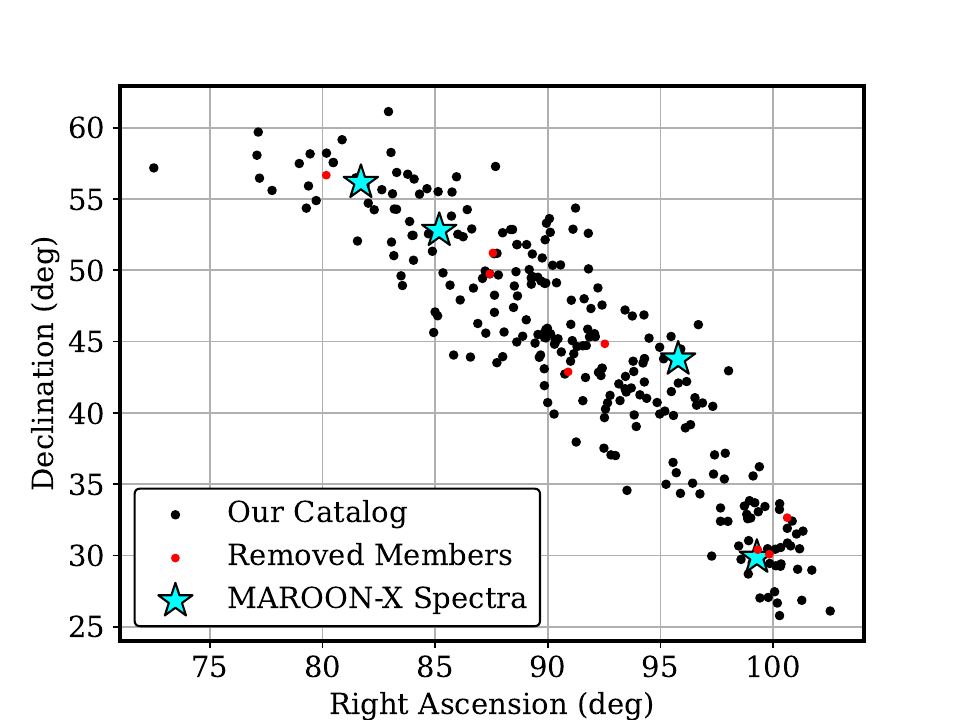}
   \caption{The sky position of OCSN-49, with right ascension vs declination plotted for members identified by the \cite{HuntCatalog} catalog. The red markers indicate stars we culled from this catalog and the black markers indicate stars included in our membership sample. The four stars for which we obtained spectra are represented by cyan markers.}\label{position}
\end{center}
\end{figure}

\begin{deluxetable*}{lccrcDD}
\centerwidetable 
\tablecolumns{7}
\tablewidth{0pt}
\decimals
\tablecaption{Gemini-N/MAROON-X Observations and Data \label{tab:observations}}
\tablehead{
    \colhead{Gaia DR3 ID}&
    \colhead{}&
    \colhead{Short ID}&
    \colhead{}&
    \colhead{\gaia $G$ (mag)}&
    \twocolhead{Integration Time (s)} &
    \twocolhead{[$S/N$]\tablenotemark{a}}
}
\startdata
263685776222367104 && GDR3-7104 && 10.95 & 700 & 100 \\
3434892295315395456 && GDR3-5456 && 11.74  &  1500 & 98 \\
962092916487309056 && GDR3-9056 && 10.95  & 700 & 103 \\
267657620243415552 && GDR3-5552 && 11.17  & 900  & 106 \\
\enddata
\tablenotetext{a}{Poisson S/N measured in the $\lambda 6700$ region.}
\end{deluxetable*}

Many of the kinematically coherent structures in the solar neighborhood uncovered by Gaia are like Theia 456: diffuse and ripe for follow-up study, both to confirm their coeval nature and further test our assumptions about star formation and the process through which stars disperse into the field. We focus on one such structure, OCSN-49, discovered by \cite{Qin_2023} and \cite{HuntCatalog}. With 265 members (per the Hunt Catalog), OCSN-49 spans almost 130 pc and 30$^{\circ}$ across the sky. In Section \ref{characteristics} we refine the OCSN-49 catalog to secure an accurate dynamical sample, and present and analyze high-resolution spectroscopy from Gemini-N/MAROON-X for four stars spanning the extent of the stream. We also measure rotation periods for 110 of its members using ground- and space-based photometric light curves. We estimate the age of OCSN-49 via isochrone fitting, lithium depletion, and gyrochronology in Section \ref{stellar ages}, and compare these with the dynamical age we derive in Section \ref{dynamical age}. In Section \ref{age discrepancy}, we note a discrepancy between the ages derived via stellar evolution and our dynamical age, and explore a scenario in which OCSN-49 was disrupted by an interaction with a giant molecular cloud (GMC). We provide some concluding thoughts in Section \ref{conclusion}.

\section{OCSN-49 CHARACTERISTICS} \label{characteristics}
\subsection{Membership} \label{membership}

To confirm the kinematic coherence of OCSN-49, we compare the positions, proper motions, and radial velocities of a random selection of stars with similar positions and parallaxes to the 265 OCSN-49 stars identified by \cite{Qin_2023} and \cite{HuntCatalog}. Background stars are selected with the following constraints: 
right ascension of 70--105$^{\circ}$, declination of 25--65$^{\circ}$, and parallax between 3.85--5.12~mas, the range in parallax spanned by OCSN-49. In Figure~\ref{background}, we display these background stars (purple) and OCSN-49 stars (black) in position and proper motion space, and include density histograms for the two populations in proper motion and radial velocity space. While OCSN-49 does not particularly stand out against the background in position, Figure \ref{background} presents a clear overdensity of stars in both proper motion and radial velocity space, visually demonstrating the stream's kinematic consistency.

In the rightmost panel in Figure \ref{background}, there are some identified OCSN-49 members with radial velocities discrepant from the median. The search algorithm used by \cite{HuntCatalog}, \texttt{HDBSCAN} \citep{HB2, HB1}, identifies overdensities in position, proper motion, and parallax, but crucially does not include radial velocity, making this last phase space dimension an independent metric for determining membership. Because OCSN-49 is quite close at a distance of $\sim$ 215 pc, most \gaia radial velocity uncertainties are $\lesssim$ 1 km s$^{-1}$, but in some cases may be higher depending on the particular star and its magnitude. We show all available \gaia radial velocities, and their uncertainties, as a function of right ascension (used as a proxy for the extent along the stellar stream) for OCSN-49 members in Figure \ref{sample}. These radial velocities are typically within a few km s$^{-1}$ of the median, far more consistent than the random sample of Milky Way background stars shown in Figure \ref{background}, which exhibit an RV dispersion of $\sim$ 36 km s$^{-1}$. This consistency suggests OCSN-49 indeed comprises a coherent structure. Six of these stars have discrepant radial velocities (indicated in red), falling more than 10 km s$^{-1}$ away from the median, and do not appear to be astrometric binaries (as indicated by their relatively low Gaia RUWE values, which are all less than 1.4). Although they may nevertheless be spectroscopic binaries, we remove these six stars, which lowers the dispersion in radial velocities from 10 km~s$^{-1}$ to 3.4 km~s$^{-1}$. Additionally, we remove two stars on the basis of proper motion in declination; the standard deviation in $\mu_{\delta}$ for all OCSN-49 stars is $\sigma_{\mu_{\delta}} = 0.65$ mas/yr, and these culled stars sit more than 3$\sigma_{\mu_{\delta}}$ away from the median. This gives us an updated membership catalog containing 257 stars which we use throughout the remainder of this work. Figure \ref{position} displays the position space of our updated catalog in black and the culled members in red.
 
\subsection{Chemical Compositions}\label{chemical compositions}

\subsubsection{Observations}\label{observations}

\begin{deluxetable*}{lDDDDcccc}
\centerwidetable 
\tablecolumns{9}
\tablewidth{0pt}
\tablecaption{Linelist and Equivalent Widths \label{tab:linelist}}
\decimals
\tablehead{
    \colhead{}&
    \twocolhead{$\lambda$}&
    \twocolhead{$\chi$}&
    \twocolhead{}&
    \multicolumn{6}{c}{Equivalent Widths (\AA)} \\
    \cline{10-13}
    \colhead{Species}&
    \twocolhead{(\AA)}&
    \twocolhead{(eV)}&
    \twocolhead{$\log gf$}&
    \twocolhead{Sun}&    
    \colhead{GDR3-7104}&
    \colhead{GDR3-5456}&
    \colhead{GDR3-9056}&
    \colhead{GDR3-5552}
    }
\startdata  
\ion{C}{1}    & 5052.167  & 7.69 & -1.30 & 29.2 &  38.9 & \nodata &  29.4 & \nodata \\
		   & 5380.337  & 7.69 & -1.61 & 18.2 &  22.7 &  20.8 &  23.8 &  20.2  \\
              & 7113.179  & 8.65 & -0.76 & 21.9 & 24.9 & \nodata &  24.9 &  21.1  \\
\ion{O}{1}    & 7771.94   & 9.15 &  0.36 & 68.6 & 99.6 &  81.0 &  98.5 &  83.9  \\
   		 & 7774.17   & 9.15 &  0.22 & 85.4 &  69.3 &  85.2 &  70.2 & \nodata  \\ 
              & 7775.39   & 9.15 &  0.00 & 69.1 &  54.6 &  66.5 &  54.3 & \nodata \\ 
\ion{Na}{1}   & 6154.226  & 2.10 & -1.56 & 37.3 & \nodata &  26.0 & \nodata &  24.1  \\  
              & 6160.747  & 2.10 & -1.26 & 54.3 & 39.8 &  41.5 &  32.6 &  35.6  \\  
\ion{Mg}{1}  	& 5711.088  & 4.35 & -1.83 & 103.2 & 89.2 &  92.5 &  84.2 &  88.0  \\
\ion{Al}{1}  	& 7362.296  & 4.02 & -0.74 & 42.1 & \nodata & \nodata & \nodata &  27.0 
\enddata
\tablecomments{This table is available in its entirety in machine-readable form.}
\end{deluxetable*}

\begin{figure}
\begin{center}

   \includegraphics[width=1.0\columnwidth]{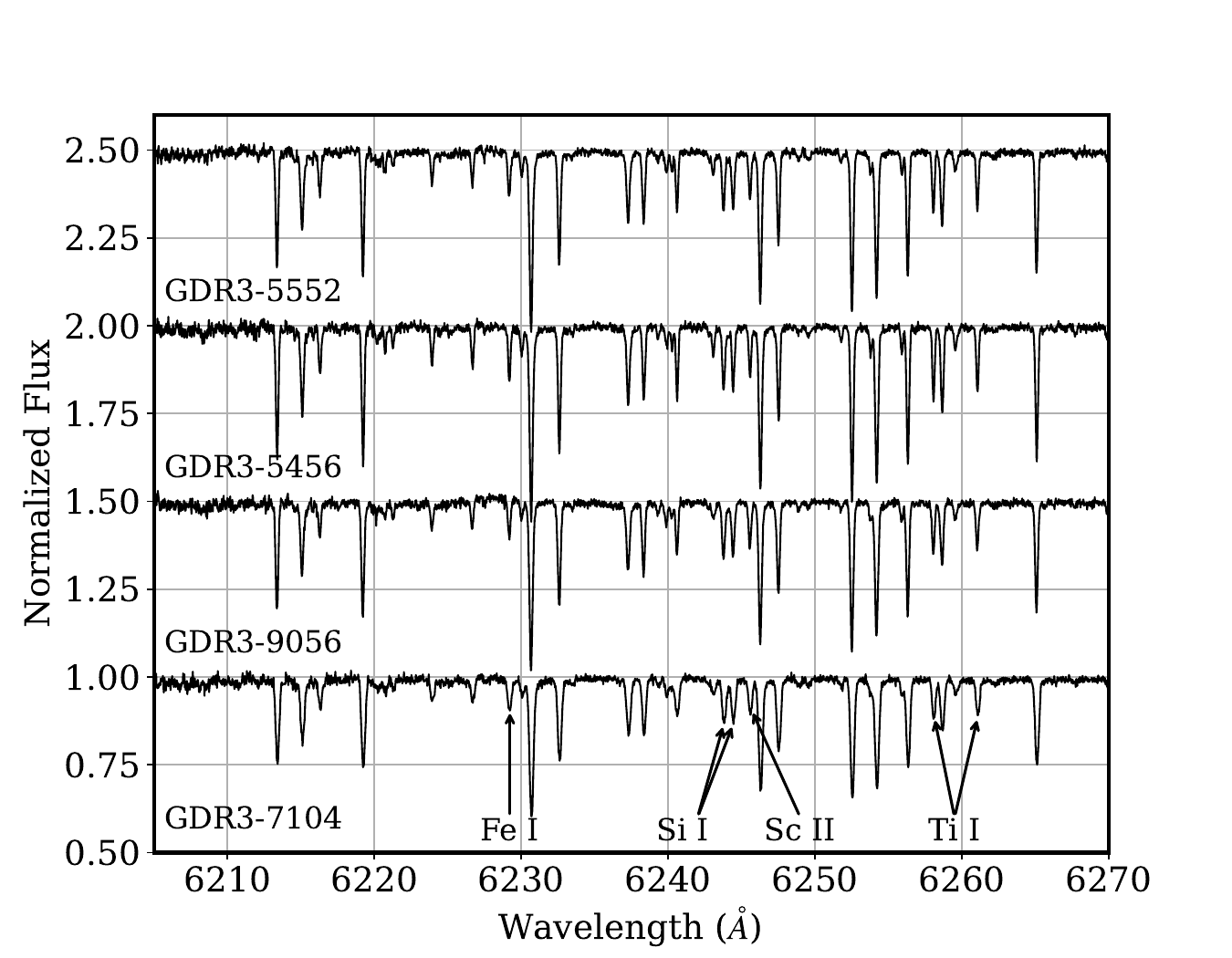}
   \caption{Sample MAROON-X spectra of the OCSN-49 stars. A sample of the lines that have been measured are indicated. Each spectra is vertically offset by 0.5 for visual clarity.}\label{spectra}
\end{center}
\end{figure}

We obtained high-resolution spectroscopy of four solar-like stars that span the length of OCSN-49, as shown in Figure \ref{position}, in order to further investigate the coeval nature of the structure, and possibly identify chemical gradients across its extent. The spectra were obtained with the MAROON-X high-resolution fiber-fed echelle spectrograph \citep{MaroonX1, MaroonX3, MaroonX2} on the 8.1-m Gemini-North telescope in Hawaii as a part of the Gemini-N Fast Turnaround program (GN-2023B-FT-201; PI:~S.~Schuler). MAROON-X is bench-mounted and highly stable, and provides a wavelength coverage of 5000--9200~{\AA} within a blue and red arm, at a spectral resolving power of $R \thickapprox 85,000$. The data were reduced by the MAROON-X support team at the University of Chicago using the observatory's custom \texttt{python3} pipeline \citep{MaroonX2}. The reduced spectra have a Poisson signal-to-noise ratio ($S/N$) of $\approx 100$ in the $\lambda 6700$ region. A solar reflection spectrum was kindly provided by A. Seifahrt, taken under observing program GN-2022A-Q-227, that included a five-minute integration on Vesta. The spectrum is of very good quality (peak $S/N$ in each order $>$ 400). Sample spectra are shown in Figure \ref{spectra}. Table \ref{tab:observations} contains a summary of the OCSN-49 observations. 

\subsubsection{Stellar Parameters, Abundances, and Errors}\label{stellar parameters}
Utilizing an equivalent width (EW) analysis, we derive for each star abundances relative to solar ([X/H]\footnote{We use the standard bracket notation to denote abundances relative, in the present case, to solar values, e.g., $\mathrm{[Fe/H]} = \log_{10} [N\mathrm{(Fe)}/N\mathrm{(H)}]_{\star} - \log_{10} [N\mathrm{(Fe)}/N\mathrm{(H)}]_{\sun}$, with $\log_{10} N\mathrm{(H)} = 12.0$.}) of up to 15 elements: C, O, Na, Mg, Al, Si, K, Ca, Sc, Ti, V, Cr, Mn, Fe, Ni. EW measurements of absorption lines, which were drawn from the linelist of \citet{2011ApJ...732...55S}, are made using the interactive Python code \texttt{XSpect-EW} \citep{XSpectEW}.  This code normalizes each order, adjusts the wavelength of the orders to align with the rest frame, fits a Gaussian profile to each line, and allows the user to modify specified parameters to adjust the fits. Each fit was visually inspected to ensure that the best possible measurements were made. The EW measurements for all the analyzed lines are given in Table \ref{tab:linelist}, as well as the wavelengths ($\lambda$), lower excitation potentials ($\chi$), and oscillator strengths ($\log gf$) of the lines.

\begin{deluxetable*}{lcccccccc}[t]
\centerwidetable
\tablecolumns{10}       
\tablewidth{0pt}   
\tablecaption{Stellar Parameters and Abundances \label{tab:abund}}             
\tablehead{                                       \colhead{}&                                   \colhead{}&
    \colhead{GDR3-7104}&                      \colhead{}&                               \colhead{GDR3-5456}&                      \colhead{}&                               \colhead{GDR3-9056}&                      \colhead{}&                               \colhead{GDR3-5552}                      }   
\startdata                                                                        
$T_{\text{eff}}$ (K)        && 6126$_{-74}^{+79}$ && 5948$_{-53}^{+63}$ && 6262$_{-77}^{+69}$ && 6074$_{-46}^{+45}$ \\                               
$\log g$ (cgs)		       && 4.43 $\pm$ 0.16    && 4.57 $\pm$ 0.10     && 4.74 $\pm$ 0.15     &&  4.63 $\pm$ 0.08   \\
$\xi$ (km s$^{-1}$)  && 1.99 $\pm$ 0.17    && 1.50 $\pm$ 0.10 		&& 2.02 $\pm$ 0.18     &&  1.63 $\pm$ 0.08   \\
{[}Fe/H{]}$_{m}$    		 && -0.001 $\pm$ 0.05  && 0.01 $\pm$ 0.03		  && 0.01 $\pm$ 0.05     && -0.02 $\pm$ 0.03   \\
$A(\mathrm{Li})$ (dex) && 3.03 && 2.76 && 3.12 && 2.86 \\
{[}\ion{Fe}{1}/H{]}     && -0.002 $\pm$ 0.05  && 0.01 $\pm$ 0.04     && 0.01 $\pm$ 0.05     && -0.02 $\pm$ 0.03   \\ 
{[}\ion{Fe}{2}/H{]}     && -0.003 $\pm$ 0.07  && 0.01 $\pm$ 0.04     && 0.01 $\pm$ 0.05     && -0.02 $\pm$ 0.03   \\  
{[}C/H{]}         && -0.09 $\pm$ 0.05   &&  0.02 $\pm$ 0.02		&& -0.12 $\pm$ 0.06    && -0.10 $\pm$ 0.07   \\  
{[}O/H{]}  	     && 0.02 $\pm$ 0.04    &&  0.02 $\pm$ 0.03		&& 0.03 $\pm$  0.04    && -0.06 $\pm$ 0.03   \\
{[}Na/H{]}  	     && -0.07 $\pm$ 0.03   && -0.12 $\pm$ 0.02		&& -0.14 $\pm$ 0.03    && -0.14 $\pm$ 0.03   \\
{[}Mg/H{]}  	     && -0.08 $\pm$ 0.04   && -0.08 $\pm$ 0.02		&& -0.11 $\pm$ 0.04    && -0.10 $\pm$ 0.02   \\
{[}Al/H{]}  	     && \nodata        		 && \nodata							&& \nodata             && -0.16 $\pm$ 0.01   \\
{[}Si/H{]}  	     && 0.02 $\pm$ 0.03    && -0.05	$\pm$ 0.02		&& -0.03 $\pm$ 0.03    && -0.06 $\pm$ 0.02   \\
{[}K/H{]}  	     && 0.15 $\pm$ 0.07    &&  0.09 $\pm$ 0.03		&& 0.05  $\pm$ 0.06    &&  0.07 $\pm$ 0.03   \\
{[}Ca/H{]}  	     && 0.05 $\pm$ 0.05    &&  0.03 $\pm$ 0.03		&& 0.03  $\pm$ 0.05    &&  0.03 $\pm$ 0.04   \\  
{[}\ion{Sc}{2}/H{]}  	 && -0.05 $\pm$ 0.10   && -0.02	$\pm$ 0.07		&& 0.02  $\pm$ 0.07    && -0.05 $\pm$ 0.05   \\
{[}\ion{Ti}{1}/H{]}  	 && -0.04 $\pm$ 0.08   && -0.01	$\pm$ 0.05		&& 0.04  $\pm$ 0.07    && -0.01 $\pm$ 0.04   \\
{[}\ion{Ti}{2}/H{]}     && -0.15 $\pm$ 0.08   && -0.02	$\pm$ 0.06		&& 0.06  $\pm$ 0.10    &&  0.05 $\pm$ 0.05   \\
{[}V/H{]}        && -0.07 $\pm$ 0.07   && -0.05 $\pm$ 0.06 	  && 0.01  $\pm$ 0.06    && -0.01 $\pm$ 0.05   \\
{[}Cr/H{]}   		 && -0.03 $\pm$ 0.05   && -0.02 $\pm$ 0.04   	&& -0.02 $\pm$ 0.04    && -0.01 $\pm$ 0.03   \\
{[}Mn/H{]}        && \nodata        		 && -0.14 $\pm$ 0.06	  && -0.15 $\pm$ 0.05    && -0.11 $\pm$ 0.02   \\
{[}Ni/H{]}        && -0.07 $\pm$ 0.06   && -0.06	$\pm$	0.04	  && -0.02 $\pm$ 0.06    && -0.05 $\pm$ 0.03   \\
{[}Cu/H{]}        && -0.43 $\pm$ 0.08   && -0.38	$\pm$	0.05		&& -0.42 $\pm$ 0.07    && -0.42 $\pm$ 0.04   \\   
\enddata         
\end{deluxetable*}  

Stellar parameters and abundances for each star are derived using the newly developed Python code \texttt{SPAE} (Stellar Parameters, Abundances, and Errors)\footnote{\url{https://github.com/simon-schuler/SPAE}}, outlined in \cite{Schuler}, which uses a Bayesian method to self-consistently generate uncertainties from the stellar atmosphere solutions when determining individual abundances. \texttt{SPAE} utilizes \texttt{MOOG} \citep[MOOGSILENT version;][]{Moog},\footnote{\url{https://www.as.utexas.edu/\~chris/moog.html}} a local thermodynamic equilibrium (LTE) plane-parallel spectral analysis code, and Kurucz model atmosphere grids \citep{Kurucz} to determine the abundances of elements in a user-provided linelist. For the purposes of this paper, we define {[}Fe/H{]} as the Fe abundance of a star relative to the solar Fe abundance and {[}Fe/H{]}$_{m}$ as the metallicity, determined by considering both \ion{Fe}{1} and \ion{Fe}{2} abundances. The Fe abundance is an output of \texttt{SPAE}, and the metallicity is used as an input parameter for a model atmosphere. 

\texttt{SPAE} starts with an adjustable initial parameter grid step consisting of an effective temperature $T_{\text{eff}}$, surface gravity $\log {g}$, metallicity {[}Fe/H{]}$_{m}$, and microturbulent velocity $\xi$. The code then interpolates the Kurucz grids for the given input parameters to produce a model atmosphere and then passes that, along with a linelist containing the EW measurements and atomic data, to \texttt{MOOG} to derive the abundances. The \ion{Fe}{1} and \ion{Fe}{2} abundances are read and used by \texttt{SPAE} to create a likelihood function formulated through the minimization of the variance between the derived \ion{Fe}{1} and \ion{Fe}{2} abundances and the model metallicity, {[}Fe/H{]}$_{m}$ \citep[see ][for details]{Schuler}. 

\begin{figure*}
\begin{center}
   \includegraphics[width=1.0\textwidth]{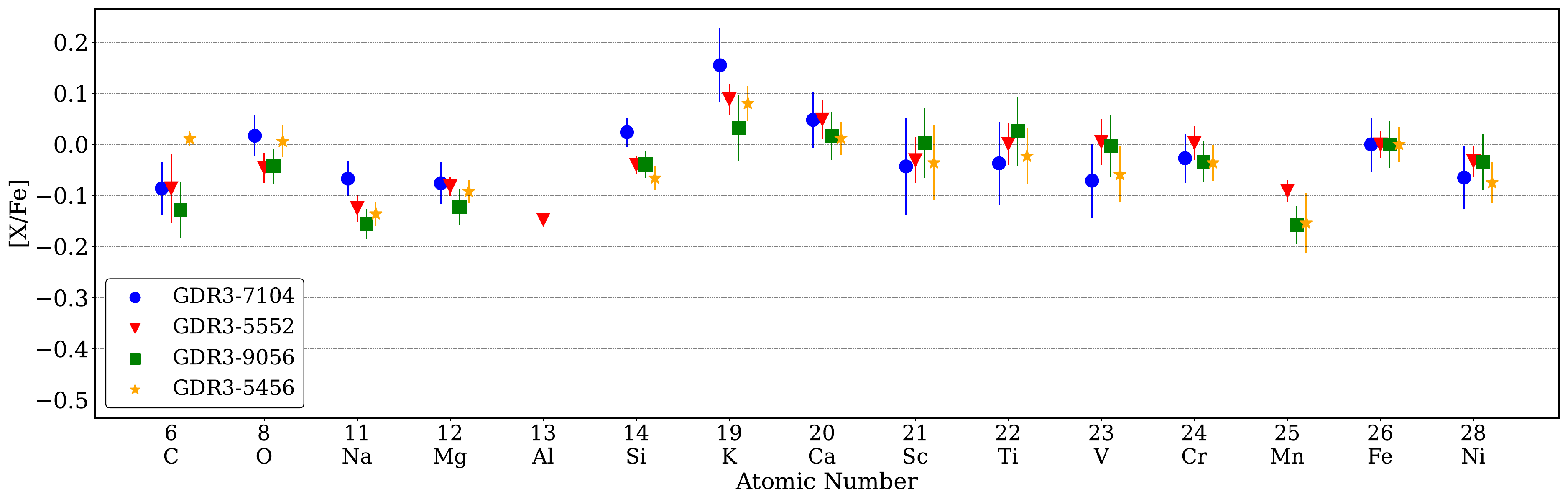}
   \caption{Element versus the median abundances and 1-$\sigma$ confidence intervals from the ensemble solution relative to Fe. The four stars are distinguished by their different markers. The consistency in the precisely measured abundances across over $\simeq15$ separate elements strongly supports the coeval interpretation of OCSN-49.}\label{abundance}
\end{center}
\end{figure*}

\begin{figure}
   \includegraphics[width=1.0\columnwidth]{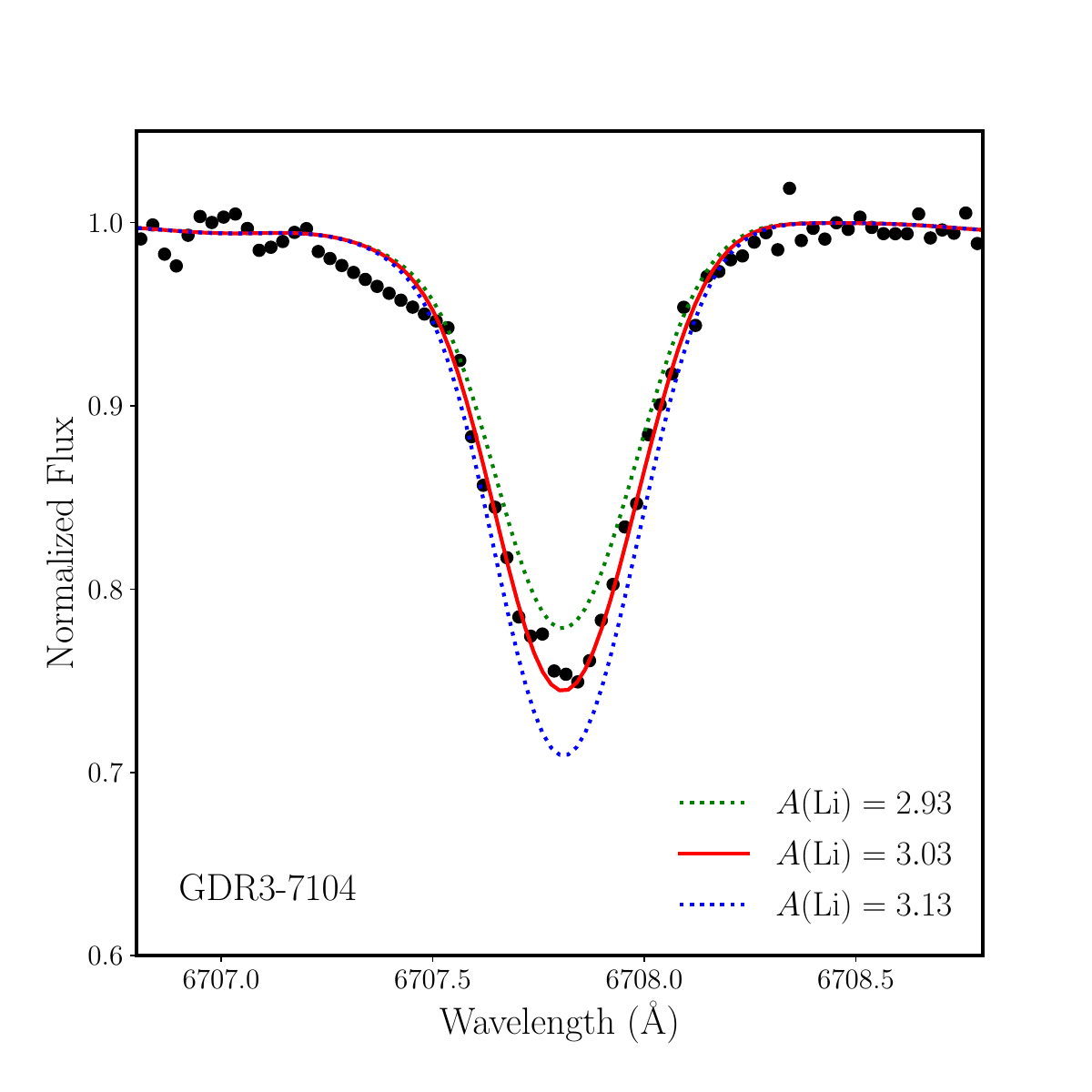}
   \caption{Example abundance synthesis of the $\lambda 6707$ Li line for GDR3-7104. Black points are the observed data, and the solid red line is the best-fit synthetic spectrum characterized by an abundance of $A(\mathrm{Li}) = 3.03$. Blue and green dotted lines are synthetic spectra with $\pm 0.10$ dex of the best-fit Li abundance. }\label{li_syn}
\end{figure}

Utilizing \texttt{emcee} \citep{emcee}---an implementation of an affine-invariant ensemble sampler MCMC algorithm---the parameter space is sampled, and the ensemble solution is generated by initializing 40 walkers in an $N$ ball around solar values ($T_{\text{eff}}=5777 ~\rm K$, $\log {g}=4.44$, and $\xi=1.38$ km s$^{-1}$) and iterating for 1,000 steps. This results in $4 \times 10^4$ separate combinations of parameters that are used to construct new model atmospheres that are passed on to \texttt{MOOG} for abundance derivations. The abundances of all elements are derived during each iteration of the MCMC algorithm. This approach, which has been tested by \citet{Schuler}, allows us to self-consistently propagate uncertainties in the stars' structural parameters in the calculation of elemental abundance uncertainties. We provide the median and 1-$\sigma$ confidence intervals of the ensemble solutions for the derived $T_{\text{eff}}$, $\log {g}$, {[}Fe/H{]}$_{m}$, $\xi$, and elemental abundances for each star in Table \ref{tab:abund}.

The metallicities of the four OCSN-49 stars are found to be closely consistent with the solar value. In Figure \ref{abundance}, we show the elemental abundances relative to Fe, plotted as a function of atomic number. Not only are nearly all of the measured abundances within 0.1 dex of the solar value, but variations in the star-to-star abundances of all elements analyzed are in excellent agreement with each other within the observational uncertainties. Although we have only measured four stars, the consistent elemental abundances of these stars suggest that the stream itself is a chemically homogeneous structure. We note that we were only able to measure Al for a single star, GDR3-5552. Finally, we point out that the abundance consistency implies there is no indication of any abundance gradient across the stellar stream's extent.

\subsubsection{Lithium Abundances}\label{lithium abundances}
Lithium abundances, $A(\mathrm{Li})$, are typically derived using spectral synthesis of the $\lambda 6707$ \ion{Li}{1} line. The line is blended with a \ion{Fe}{1} line that can contribute to the absorption of the $\lambda 6707$ line, particularly in more metal-rich stars, so synthesis is preferred to a standard EW analysis. The median solutions for the stellar parameters of each star from \texttt{SPAE} are used to interpolate Kurucz model atmospheres, and \texttt{MOOG} is used to produce the synthesized spectra, making use of the linelist from \citet{2022MNRAS.513.5387S} based on the original work of \citet{1997AJ....113.1871K}.

To derive $A(\mathrm{Li})$, synthesized spectra are fit to the  $\lambda 6707$ line in the observed spectrum of each star, and the best fit is determined by eye. Rotational and instrumental broadening of the synthetic spectrum is modeled using a Gaussian smoothing function within \texttt{MOOG} and is constrained by fitting nearby \ion{Fe}{1} lines in the same spectral region. We show an example of the $\lambda 6707$ line fits for GDR3-7104 in Figure \ref{li_syn}. The observed spectrum along with the best fit and syntheses with $\pm0.10 \; \mathrm{dex}$ of the best-fit Li abundance are shown. We note that the abnormal data on the blue shoulder of this particular line is an artifact from the cosmic ray removal procedure. Although we do not provide uncertainties on $A(\mathrm{Li})$ in Table~\ref{tab:abund}, comparison with the $\pm0.10 \; \mathrm{dex}$ lines in Figure~\ref{li_syn} suggests the uncertainty is significantly smaller than 0.1 dex.

\subsection{Rotation Periods}\label{rotation periods}
We used time series imaging data from ground- and space-based observatories to measure rotation periods for 110 members of OCSN-49. 

The Transiting Exoplanet Survey Satellite \citep[TESS;][]{Ricker2015} observed members of OCSN-49 during Cycles 2 ($N$=114 stars), 4 ($N$=26 stars), 5 ($N$=115 stars), and 6 ($N$=124 stars), typically for one sector per cycle (each sector spans $\approx$27 days). We downloaded 40$\times$40 pixel cutouts from the Full-Frame Images (FFIs) using \texttt{TESScut} \citep{TESScut}, then extracted light curves using the causal pixel modeling code \texttt{unpopular} \citep{Hattori2022}. The observing cadence varies across the cycles from as long as 30~min (Cycle 2) to 200~s (Cycles 5--6)---we binned the light curves down to 30~min cadences for all cycles. Next, we used Lomb--Scargle periodograms \citep{press1989} and visual inspection to classify stars as periodic and to measure periods for 103 stars out of 128 targets (80\% success rate). The multiple TESS cycles were essential for identifying and correcting half-period harmonics, where symmetric spot patterns on opposing hemispheres create a double-dipping light curve pattern resulting in an apparent period half the value of the true rotation period.

The Zwicky Transient Facility \citep[ZTF;][]{ztf} has imaged the northern sky from Palomar Observatory $\sim$nightly since March 2018. Following \citet{Curtis_2020}, we downloaded 8$\times$8 arcmin ZTF-$r$ band images and performed circular aperture photometry on our targets and nearby reference stars identified with Gaia, which we used to detrend the resulting light curves. We targeted stars with $13<G<17$ (89 stars), and inspected their light curves and Lomb--Scargle periodograms for each available season to identify periodic stars, and we measured periods for 59 stars (66\% success rate). 

For 52 of the 59 stars with ZTF periods, we also measured the period from TESS. Almost all of these overlapping periods agree to within 5\% (46 of 52 stars), and the median fractional difference is $\approx$1\%. For these stars, we report the average of the TESS and ZTF periods.

\section{Stellar Ages}\label{stellar ages}
\subsection{Isochrone Age} \label{isochrone age}

\begin{figure}
   \includegraphics[width=1.0\columnwidth]{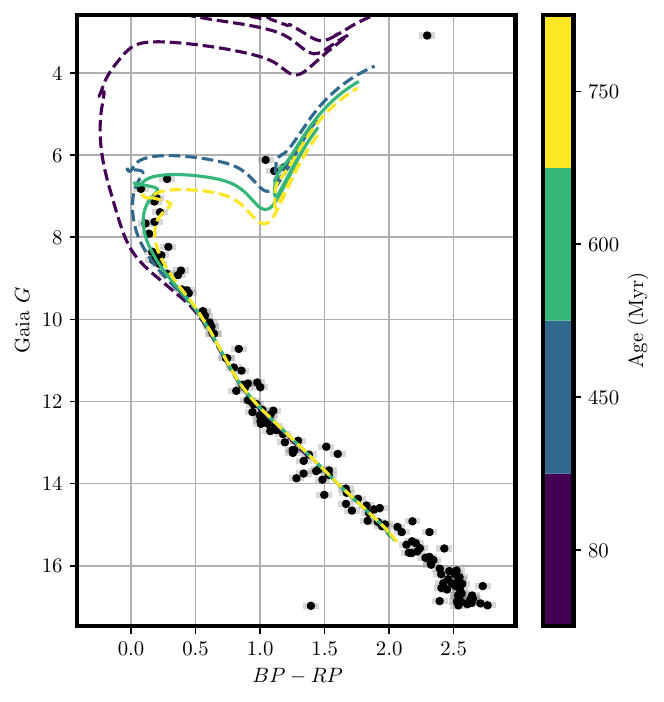}
\caption{Color--magnitude diagram of all OCSN-49 stars with \gaia $G \leq 17$ , with our best-fit isochrone (600 Myr, solid line) as well as three other isochrones at different ages. OCSN-49's color--magnitude diagram clearly indicates that its stars are inconsistent with the dynamical age ($\simeq$80 Myr) we derive in Section \ref{model}.}\label{isochrones}
\end{figure}

\begin{figure*}
\begin{center}
   \includegraphics[width=0.5\textwidth]{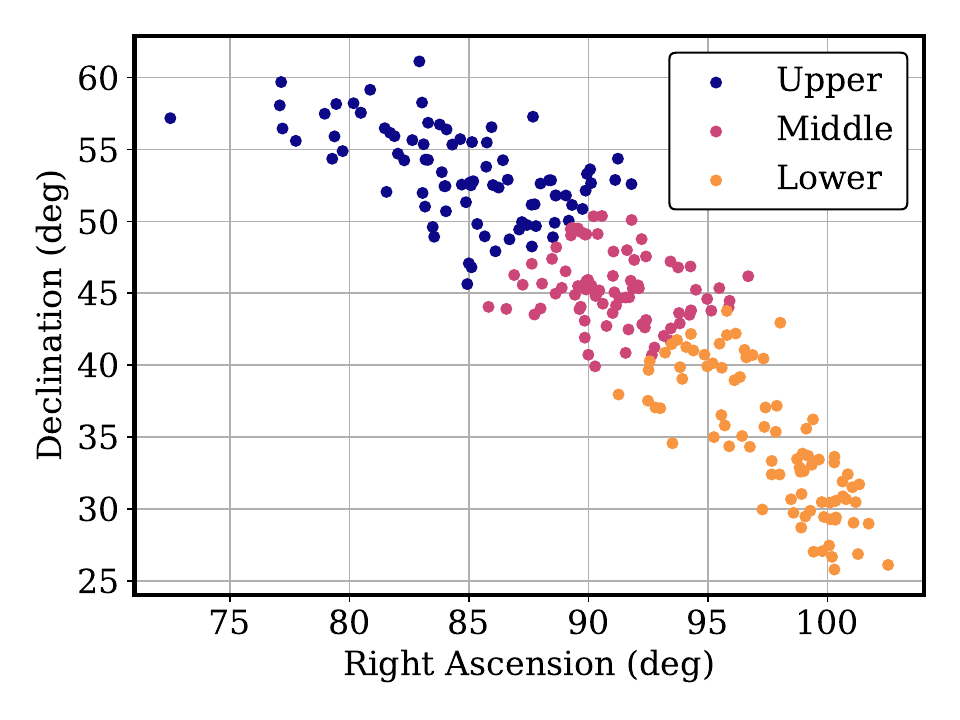}
   \includegraphics[width=1.0\textwidth]{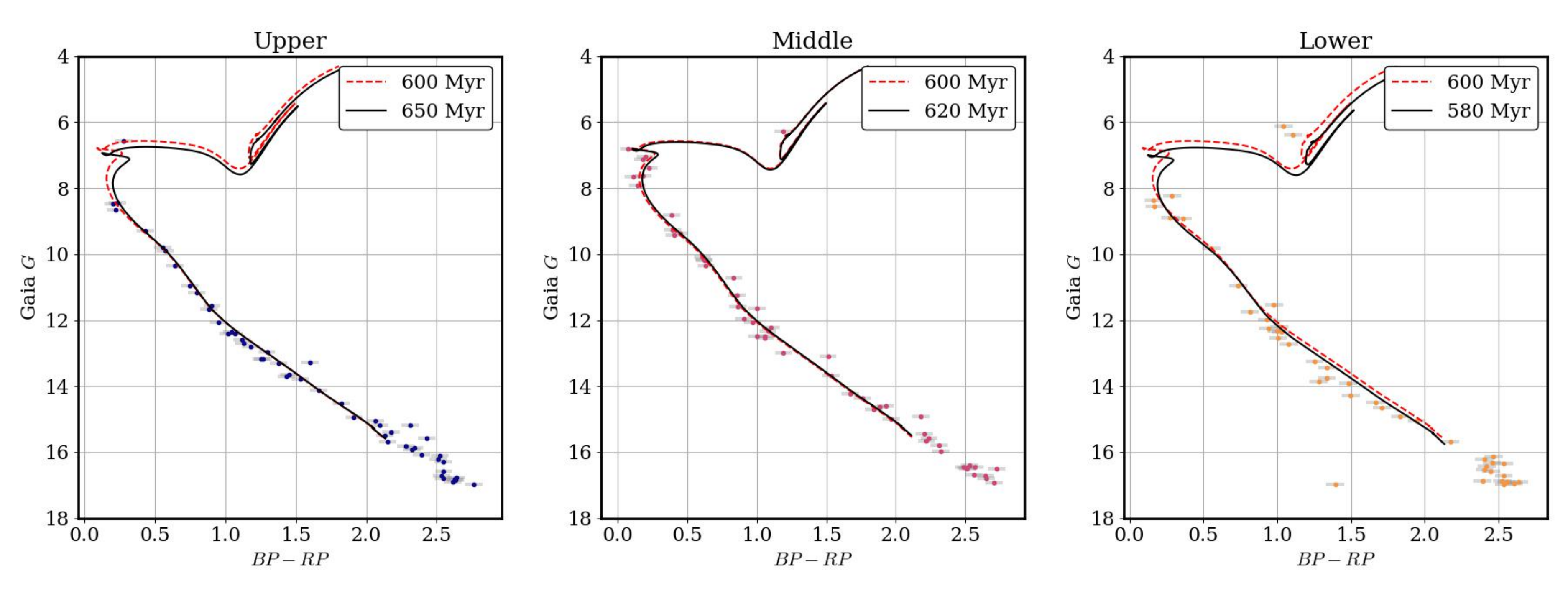}
\end{center}
\caption{\textit{Top}: Our defined upper, middle, and lower regions of the stream. \textit{Bottom}: The resulting isochrone fits to the corresponding stream
sections. For the upper section, we find an age of $\simeq$650 Myr and distance of $\simeq$210 pc. For the middle section, we find $\simeq$620 Myr and distance of $\simeq$210 pc, and for the lower section we find $\simeq$580 Myr and distance of $\simeq$230 pc. We find the age estimate of $\simeq$600 Myr (shown as a dashed red isochrone) to be consistent across the stream.}\label{lobesandisochrones}
\end{figure*}

To determine an age estimate for OCSN-49, we turn to the commonly used technique of isochrone fitting \citep[for a review, see][]{ageofstars}. We utilize \texttt{Brutus},\footnote{\url{https://github.com/joshspeagle/brutus}} a Python package that simultaneously fits age, distance, reddening, extinction, and metallicity to theoretical MIST isochrones \citep{Mist1, Mist2}. \texttt{Brutus} takes in \gaia photometric data and parallaxes (and their associated errors) for individual stars as a prior. We maximize the built-in likelihood function using the Powell method in \texttt{SciPy} \citep{Scipy}, as more sophisticated optimization routines were not necessary to determine best fit parameters for the entire stream.

Due to discrepancies between the data and the MIST model isochrones at lower masses, we only include stars with \gaia $G \leq 17$ for our fit. Additionally, based on the fits to our four stars with high-resolution spectra described in Section~\ref{stellar parameters} we fix the metallicity to be solar. \texttt{Brutus} also includes an outlier fraction in its model; we fix this fraction to be 10\%. We choose to only fit the single star sequence in our model, ignoring any binary tracks. Following these assumptions, our fit to the entire stream produces an age of $\simeq$600 Myr, and a distance of $\simeq$210 pc, consistent with the distance derived via a weighted average of $1 / \varpi$. This age is bracketed by previously reported ages for OCSN-49 of 470 Myr by \citet{HuntCatalog} and 700 Myr by \citet{Qin_2023}. 

In Figure \ref{isochrones}, we show our best fit isochrone as a solid line, along with additional isochrones at 130, 450, and 750 Myr (but the same metallicity, reddening, extinction, and distance) for comparison. OCSN-49's disperse nature, however, presents a challenge: the structure spans a range of parallaxes, from 3.85--5.12 mas, which results in a gradient of distance across the structure from 195--260 pc. To confirm that our isochrone fit to the entire stream produces a robust age estimate, we divide the stream into three sections, and separately follow the same fitting procedure with \texttt{Brutus} on each subset. In Figure \ref{lobesandisochrones}, the top panel shows how we divide the stream, such that each section has roughly the same number of stars. The bottom three panels show the resulting isochrone fits to the corresponding stream sections. For the upper section, we find an age of $\simeq$650 Myr and distance of $\simeq$210 pc. For the middle section, we find $\simeq$620 Myr and distance of $\simeq$210 pc, and for the lower section we find $\simeq$580 Myr and distance of $\simeq$230 pc. We therefore conclude that 600 Myr provides a reasonable estimate of OCSN-49's stellar age.

\subsection{Lithium} \label{lithium age}

\begin{figure*}
\begin{center}
   \includegraphics[width=1.0\textwidth]{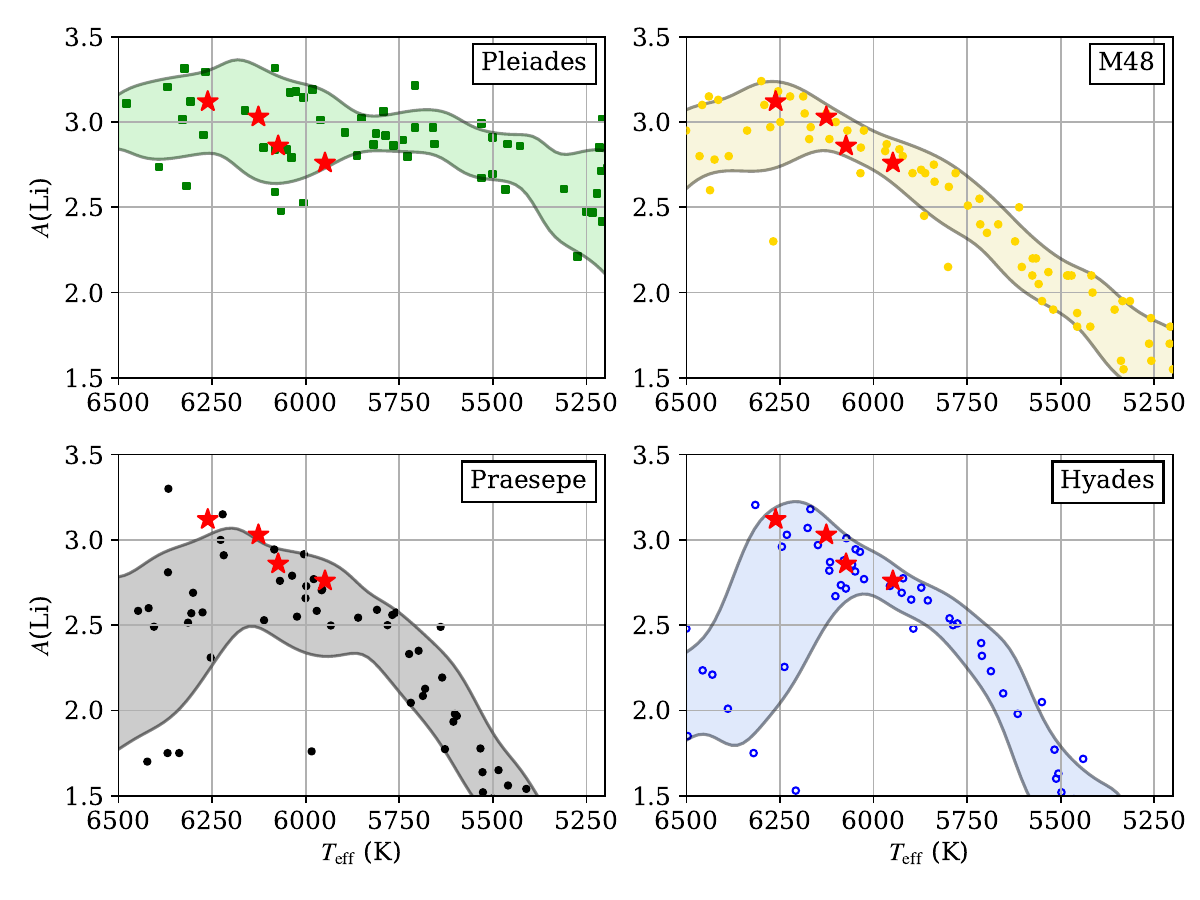}
   \caption{We compare $A$(Li) for our four OCSN-49 stars to stars in four separate clusters: the Pleiades \citep[125 Myr;][]{2018A&A...613A..63B}, M48 \citep[420 Myr;][]{2023ApJ...952...71S}, Praesepe \citep[700 Myr;][]{Douglas2019,Rampalli2021}, and the Hyades \citep[680 Myr;][]{Gossage_2018}. While our sample is potentially consistent with any of these, our likelihood ratio tests suggest M48 is significantly more consistent. }\label{fig:lithium_matching}
\end{center}
\end{figure*}

As stars evolve, their lithium content can change as a result of a complex interplay between lithium destruction due to burning at the base of the convective zone \citep{1997ApJ...482..442B} and internal mixing processes, likely modulated by stellar rotation \citep{1993AJ....106.1059S}. As a population, these effects produce rich structure in the lithium abundances of stars of different temperatures, a structure that evolves as lithium is progressively destroyed at a mass-dependent rate. Lithium can therefore be used to provide an independent age measurement of coeval stellar populations \citep{ageofstars}.

In Figure~\ref{fig:lithium_matching}, we compare the $A$(Li) and $T_{\rm eff}$ for our sample of four OCSN-49 stars observed with Gemini-N, against four well-studied open clusters with known ages: the Pleiades \citep[125 Myr;][]{2018A&A...613A..63B}, M48 \citep[420 Myr;][]{2023ApJ...952...71S}, Praesepe \citep[700 Myr;][]{Douglas2019,Rampalli2021}, and the Hyades \citep[680 Myr;][]{Gossage_2018}. While the Pleiades only shows a slight decline in $A$(Li) over the plotted range of $T_{\rm eff}$, the similarly aged Hyades and Praesepe show a significant reduction for stars colder than $\simeq$6000 K. Having an age between these two extremes, M48 shows an intermediate decline. 

While our four stars are not terribly inconsistent---at least by eye---with any of these four distributions of stars, we can nevertheless perform an initial statistical comparison. We start by generating a moving average of the spread in $A$(Li) as a function of $T_{\rm eff}$, adopting a Gaussian kernel with a variance of $(100 ~\rm{K})^2$. The colored contours in each panel display the 1$-\sigma$ limits of that moving average, assuming a Gaussian distribution in $A$(Li). We can then quantify the consistency by calculating a likelihood,

\begin{equation}
    \mathcal{L} = \prod_{i=1}^4 \mathcal{N}\left[A({\rm Li})_i; \mu(T_{\rm eff}), \sigma(T_{\rm eff})\right],
\end{equation}
where the Gaussian normal distribution $\mathcal{N}$ is evaluated for the $i$th star's $A$(Li)$_i$ using the mean $\mu$ and standard deviation $\sigma$ calculated from the kernel-weighted moving average at that star's $T_{\rm eff}$. Applying this approach to our four OCSN-49 stars, we find likelihoods of 2.8 for the Pleiades, 32.7 for M48, 0.6 for Praesepe, and 2.9 for the Hyades. We therefore find that the lithium abundances of our OCSN-49 stars are an order of magnitude more similar to M48 compared with the other clusters. While one should be cautious of over-interpretation with just four stars, this analysis suggests there is a slight statistical preference for OCSN-49 to have an age closer to $\simeq$420 Myr, somewhat younger than the isochrone age of $\simeq$600 Myr.

\begin{figure*}
\begin{center}
   \includegraphics[width=1.0\textwidth]{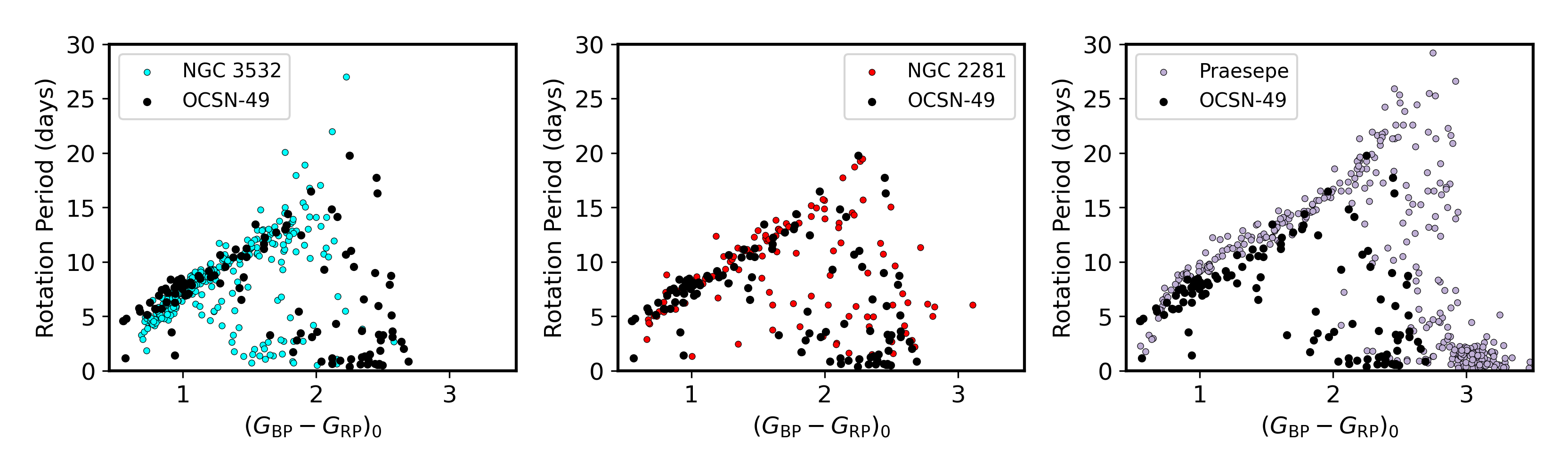}
   \caption{Color--period diagrams for OCSN-49 (black dots) versus NGC~3532 (400~Myr; left), NGC~2281 (500~Myr; middle), and Praesepe (700 Myr; right panel). The rotation period distribution for OCSN-49 is most consistent with NGC~2281, indicating that they share a common age of $\approx$500~Myr. The rotation period measurements are available as Data behind the Figure.}\label{cpd}
\end{center}
\end{figure*}

\begin{figure*}

   \includegraphics[width=1.0\textwidth]{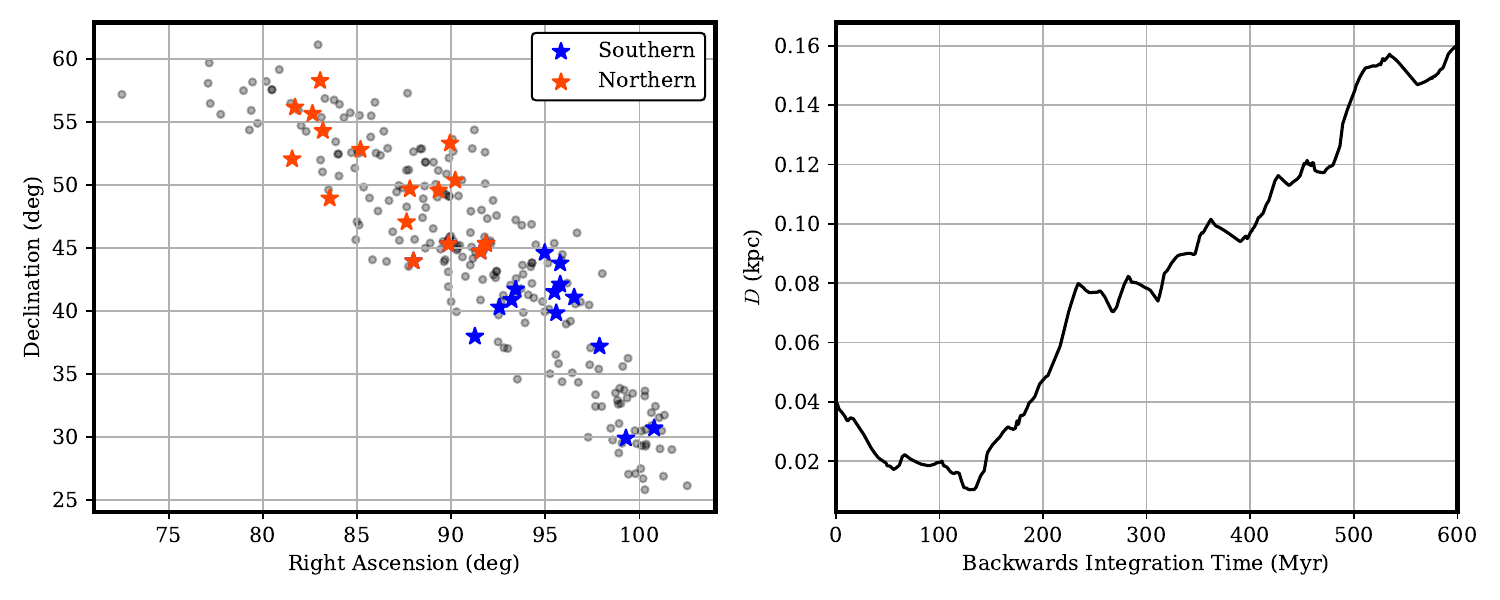}
   \caption{\textit{Left}: Right ascension versus declination of OCSN-49 members with radial velocity errors less than 1 km s$^{-1}$ in color. We divide this resulting sample in half with a slice perpendicular to the extent of the stream. We classify the resulting regions into northern (red) and southern (blue) stars. \textit{Right:} Distance between the two regions ($D$) versus backwards integration through a Milky Way potential. The two regions come within 10 pc of each other about 130 Myr ago.}\label{bwardsint}

\end{figure*}

\begin{figure*}
\begin{center}
   \includegraphics[width=1.0\textwidth]{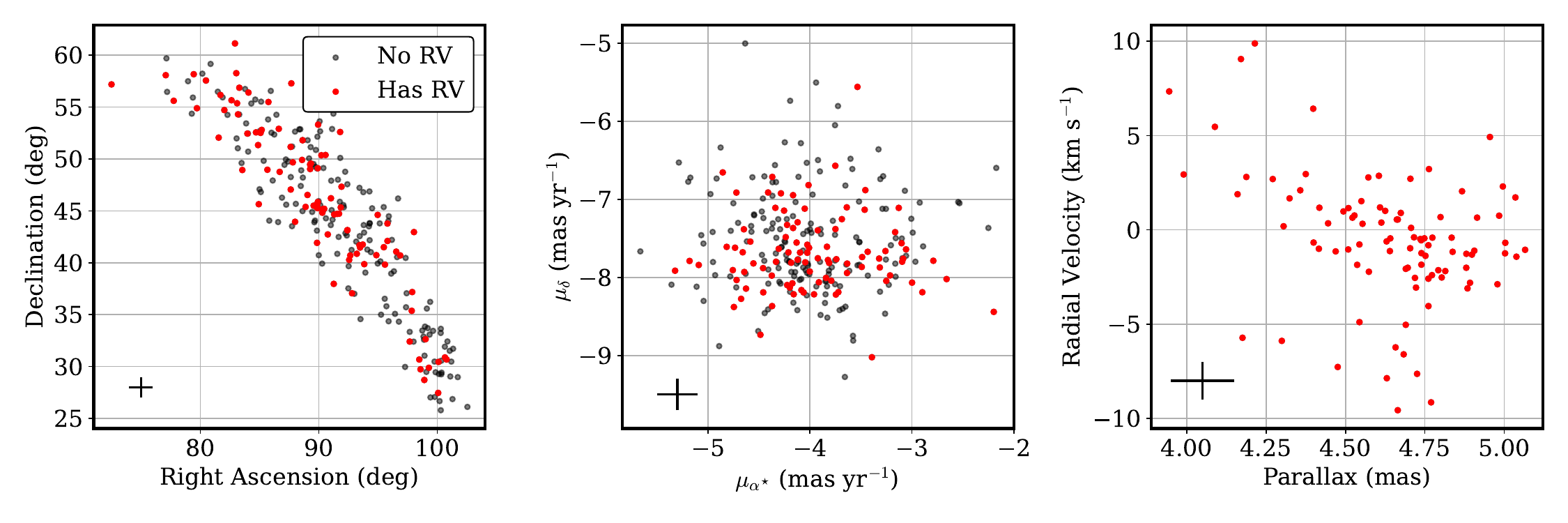}
   \caption{Position, proper motion, and parallax-radial velocity space plotted for all 257 OCSN-49 stars. Stars with Gaia radial velocities are identified in red and those without are identified in gray. Only members with measured Gaia radial velocities are plotted in parallax-radial velocity space. Error bars are chosen to apply to all stars such that they are overlapping but do not overshoot the width of the distributions, shown in the bottom left corner of each panel. We arrive at errors of 1$^{\circ}$ for position, 0.1\text{ mas} for parallax, 0.2\text{ mas yr$^{-1}$} for proper motion, and 1\text{ km s$^{-1}$} for radial velocity (for stars with Gaia radial velocity errors less than 1\text{ km s$^{-1}$}).}\label{errorbars}
\end{center}
\end{figure*}

\subsection{Gyrochronology} \label{gyrochronology}

Stars tend to be born spinning rapidly, and their spin slows over time due to magnetic braking \citep{Skumanich_1972, ageofstars}. Observations of rotation in benchmark open clusters allow for the empirical calibration of gyrochronology \citep{Barnes_2003}, which can then be applied to age-date individual field stars or clusters. Any kinematic structure that is truly composed of stars born together should have a large number of its members showing a rotation period distribution consistent with known benchmark star clusters \citep[e.g.,][]{Curtis2019_PscEri}.

In Figure \ref{cpd}, we compare our rotation periods for OCSN-49 directly to those for NGC~3532 \citep[400~Myr;][]{Fritzewski2021_3532}, 
NGC~2281 \citep[500~Myr;][]{Fritzewski2023_2281}, and
Praesepe \citep[700~Myr;][]{Douglas2019,Rampalli2021}.
In the left panel, we see that compared with OCSN-49, the bluer stars in NGC 3532 are still rapidly rotating, indicating OCSN-49 is likely older than 400 Myr. At the same time, the right panel of Figure \ref{cpd} shows that the transition from rapid to slow rotators occurs at redder colors in Praesepe compared with OCSN-49, indicating OCSN-49 is likely younger than 700 Myr. The matching distributions in the middle panel indicate that OCSN-49 likely has an age close to 500 Myr, consistent with our estimates based on isochrones and lithium. We further note that our rotation sample for OCSN-49 is large: over 80\% of our gyrochronology targets (i.e., those with $G_{\rm BP}-G_{\rm RP} > 0.6$ and $G<17$) have measured periods. The consistency of the rotational distribution of OCSN-49 in Figure~\ref{cpd} further supports our conclusion that the stars share a common origin, as non-coeval stars would be expected to produce a more scattered distribution in the color-rotation period plane.

\section{DYNAMICAL AGE} \label{dynamical age}
\subsection{Backwards Integration}\label{backwards integration}
In order to assess when the stream was most compact in position space, we utilize \texttt{gala} \citep{Gala}, a Python package designed to take in six-dimensional (6D) phase space information and integrate orbits through a Galactic potential. With the default parameters of \texttt{Astropy v4.0} \citep{Astropy}, we utilize the built-in \texttt{MilkyWayPotential}, a multi-component model potential fit to recent mass measurements of the Milky Way.

\begin{figure*}
\begin{center}
   \includegraphics[width=1.0\textwidth]{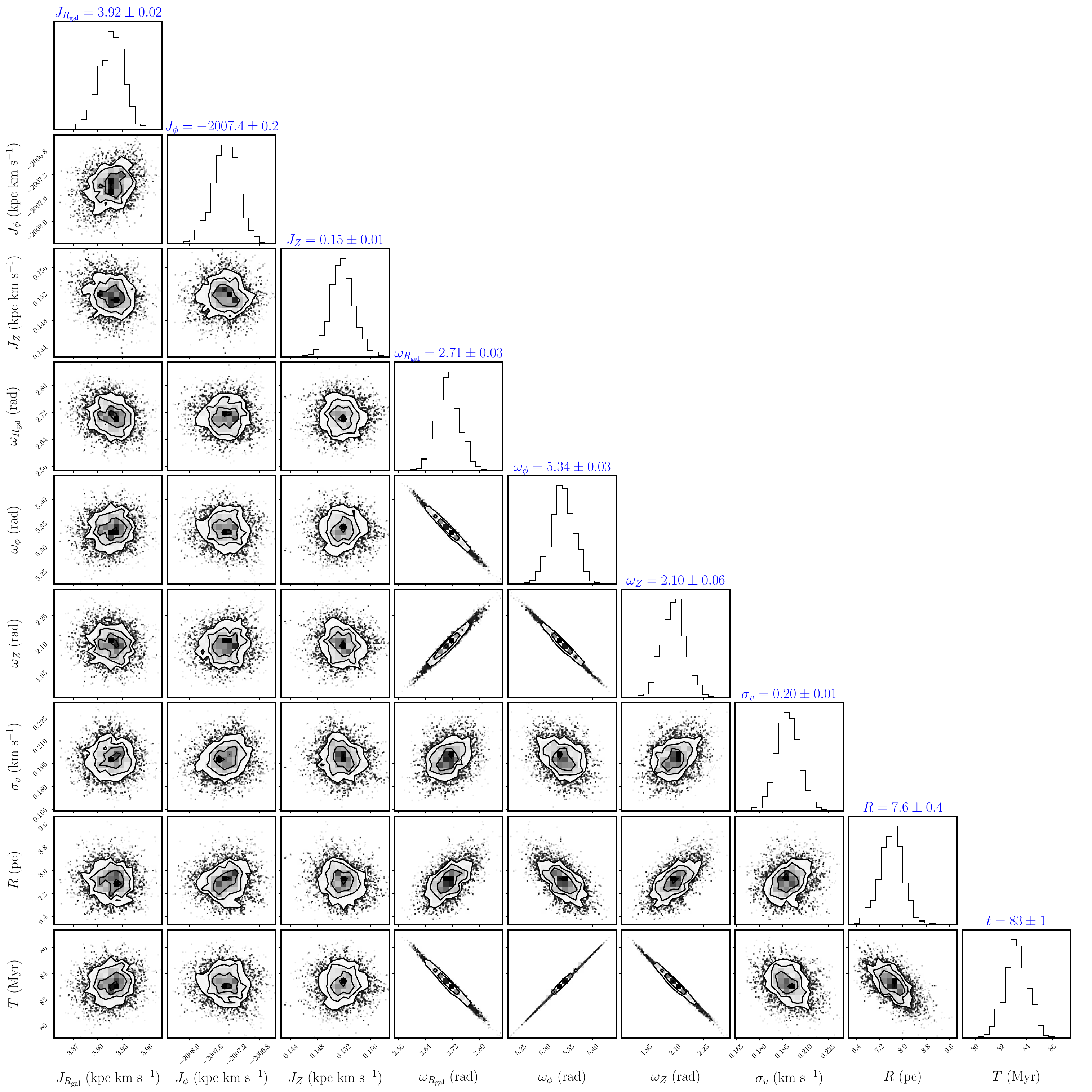}
   \caption{Resulting model parameters of OCSN-49 birth conditions ($\psi_{0, \rm opt}$) derived from our statistical model. We plot only the lowest temperature $(T = 1)$ walkers, which sample the posterior space. }\label{cornerplot}
\end{center}
\end{figure*}

To more accurately integrate orbits, we use the subset of stars in our sample with Gaia radial velocity uncertainties less than 1 km s$^{-1}$. To further reduce the effect of individual stars' measurement uncertainties, we average over many stars, dividing this resulting subsample of 29 stars into a northern and southern region, as seen in the left panel of Figure \ref{bwardsint}, where colored stars have sufficiently small uncertainties in radial velocity. To understand the evolution of these two regions, we separately evolve each star backwards in time and define $D$ as the distance in 3D space between the mean $X$, $Y$, and $Z$ positions of the northern and southern stars at each time slice. The right panel of Figure \ref{bwardsint} shows $D$ as a function of backwards integration time. The two regions start about 40 pc apart and come closest in space, separated by $\simeq$10 pc about 130 Myr ago. This small separation, approaching the size of a typical open cluster, suggests OCSN-49 originated only $\sim100$ Myr ago.

\subsection{Statistical Model Parameters and Results} \label{model}

\begin{figure*}
\begin{center}
   \includegraphics[width=1.0\textwidth]{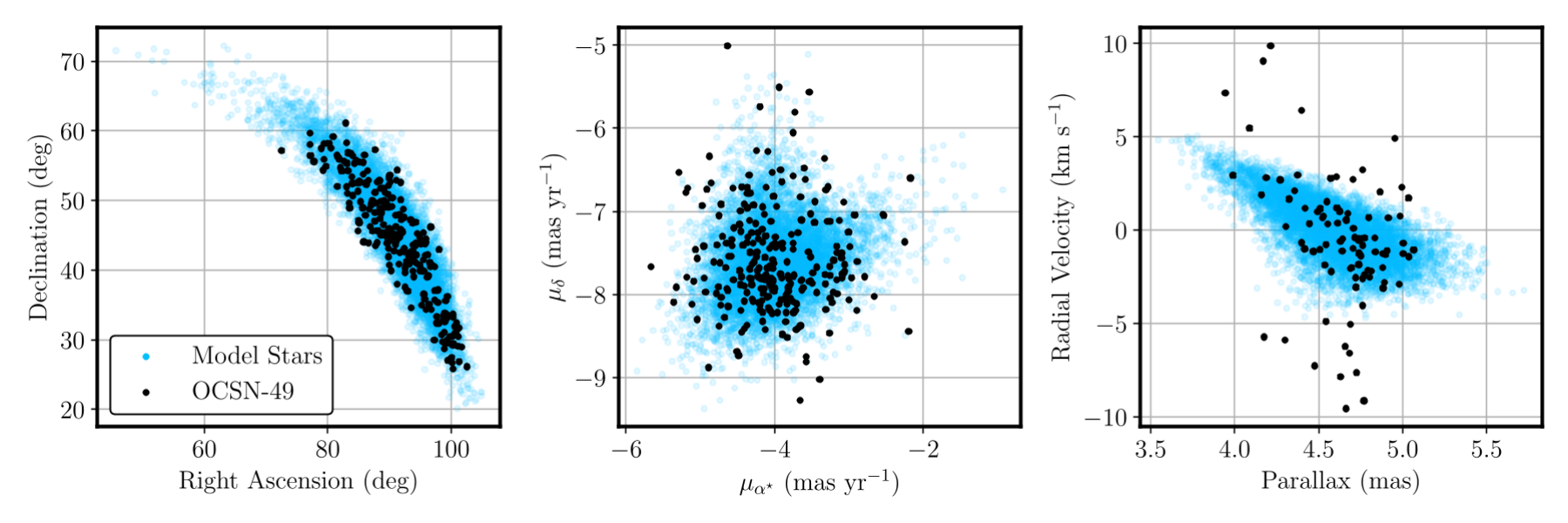}
   \caption{Forward integrated model stars with our resulting model parameters we see in Figure \ref{cornerplot} (blue), compared with OCSN-49 as we observe today (black). All 257 OCSN-49 stars are plotted in position and proper motion space, with only the 93 stars with measured Gaia radial velocities plotted in parallax-radial velocity space.}\label{posteriorcheck}
\end{center}
\end{figure*}

To derive a more robust dynamical age estimate of OCSN-49, we adapt the Bayesian statistical model outlined in \cite{Kyle}. This model approximates the progenitor cluster of OCSN-49 as an unbound sphere of stars, with a characteristic size and velocity dispersion. By integrating these stars forwards in time, we can aim to reproduce the size, shape, and location of OCSN-49 in phase space today. These stars are assumed to be tracer particles, where self-gravity from the mutual gravitational attraction of component stars is ignored. Our model parameters are defined as:
\begin{equation*}
\psi_0 = \{J_{R_{\rm gal}}, J_{\phi}, J_Z, \omega_{R_{\rm gal}}, \omega_{\phi}, \omega_Z,\sigma_v, R, t\}
\end{equation*}
where the first six parameters are action-angle coordinates in a cylindrical Galactocentric reference frame, $\sigma_v$ is an isotropic velocity dispersion, $R$ is the initial scale of the tracer cluster, and $t$ is the cluster age. We utilize action-angle coordinates to reduce covariances between sample parameters. The model is populated as a 3D Gaussian, where $R$ is the standard deviation in position in all three dimensions and $\sigma_v$ is the standard deviation in velocity in all three dimensions. We give the model the following measured observables: 
$$x^{\prime}_{f,k} = (\alpha, \delta, \varpi, \mu_{\alpha^{\star}}, \mu_{\delta}, V_{\text{rad}})$$
where $V_{\rm rad}$ is ignored for stars without Gaia radial velocities, and included for the 93 stars with radial velocities.

To find optimal model parameters, we use \texttt{ptemcee} \citep{ptemcee}, a parallel-tempered version of \texttt{emcee}. Parallel-tempering modifies the posterior by attaching an exponent to the likelihood function. In log-space, this translates to:
\begin{equation}
    \log{P(\psi_0 | \{x_f^{\prime}\}) } \propto \frac{1}{T} \log{\mathcal{L}(\{x_f^{\prime}\} | \psi_0 )} + \log{P(\psi_0)}
\end{equation}
where $P(\psi_0 | \{x_f^{\prime}\})$ is the modified posterior, $\mathcal{L}(\{x_f^{\prime}\} | \psi_0 )$ is the likelihood function, $T$ is the temperature, $P(\psi_0)$ is the prior, and $\{x_f^{\prime}\}$ refers to the set of all $K = 257$ measured observables, $x^{\prime}_{f,k}$. Our log-likelihood function is given by the first term in equation 7 in \cite{Kyle}, and we choose our prior to be flat for all nine parameters. $T = 1$ corresponds to the posterior distribution as given by Bayes Theorem, and the modified posterior distribution approaches the prior distribution as $T$ tends to infinity. 

Our isochrone fit to OCSN-49's photometry indicates an age of $\simeq$600 Myr, meaning this entire range must be explored in parameter space. This leads to a multi-modal distribution in posterior space that can be problematic for some optimizers. In \texttt{ptemcee}, walkers are assigned to different temperatures, and allowed to jump between them. Walkers at high temperatures explore the entire prior space in order to escape local maxima in the posterior distribution. Temperature swaps occur throughout the simulation; the utility of parallel-tempering occurs when walkers at high temperatures find the global maximum in the likelihood, where they then jump to lower temperatures to explore the posterior space.

PTMCMC allows for temperatures to be dynamically adjusted to achieve uniform swap acceptance ratios between adjacent temperatures in the chain \citep[see section 3 in][]{ptemcee}. We initialize 8 logarithmically spaced temperatures from $T=1$ to $T=100$, with 64 walkers each, and run our PTMCMC simulation for 4,000 steps, initializing the simulation at $t = 250$ Myr and allowing dynamic temperature adjustment. At each step, $N = 10,000$ model stars are generated at a given $\psi_0$, and action-angle coordinates are converted to physical coordinates using the \texttt{actionAngleTorus} implementation in \texttt{Galpy} \citep{galpy}. Model stars are then integrated under our model Milky Way potential for $t$ Myr. The posterior is computed by comparing the 6D phase space of each model star to the phase space of each observed OCSN-49 star (either 5D or 6D, depending on whether that observed star has a Gaia radial velocity). For more details on this calculation, see Section 4.1 in \cite{Kyle}. 

Since \gaia's uncertainties are too small for numerical convergence in a reasonable time-frame, we augment the errors to represent the broader distribution of OCSN-49, as we are not interested in reproducing the phase space of each star individually, only the distribution. We essentially adopt a kernel density estimate of our $K=257$ stars, choosing the following kernel sizes to each dimension applied uniformly: $\sigma_{\alpha} = \sigma_{\delta}=1^{\circ}$, $\sigma_{\varpi}=0.1\text{ mas}$, and $\sigma_{\mu_{\alpha^{\star}}} = \sigma_{\mu_{\delta}} = 0.2 \text{ mas yr}^{-1}$. For stars with Gaia radial velocity errors less than $1 \text{ km s}^{-1}$, we choose $\sigma_{V_{\text{rad}}} = 1 \text{ km s}^{-1}$, otherwise, we use the radial velocity error reported by Gaia. These values allow the data points to have overlapping error bars, but they do not overshoot the width of the distributions, as shown in Figure \ref{errorbars}. Additionally, we perform convergence tests to ensure that our chosen value of $N$ is large enough and our orbital integration time-step is small enough to be robust across various initial random draws.

Figure \ref{cornerplot} shows our resulting model parameters from the lowest temperature $(T = 1)$ walkers and their corresponding uncertainties, with burn-in removed. We obtain a resulting velocity dispersion of $\sigma$$_{v_{\text{opt}}}$ = 0.20 $\pm$ 0.01 \text{km s$^{-1}$} and scale of $R_{\text{opt}}$ = 7.6 $\pm$ 0.4 \text{pc}, which corresponds to a half-mass radius of 11.4 $\pm$ 0.6 \text{pc}.

Of particular note is the best-fit cluster age:
\begin{equation*}
    t_{\text{opt}} = 83 \pm 1 \text{ Myr}.
\end{equation*}
This age is in line with our initial estimates from integrating individual stellar orbits backwards in time as described in Section~\ref{backwards integration}, but highly inconsistent with the stellar age of $\simeq$600 Myr derived in Section~\ref{stellar ages}. We address this inconsistency in Section~\ref{age discrepancy}.

\subsection{Posterior Predictive Checking} \label{posterior predictive}

To assess the validity of our model parameters, and by extension the derived dynamical age, we perform posterior predictive checking to confirm that our derived model parameters reproduce the observed phase space positions of OCSN-49 stars today. We generate 10,000 model stars with the maximum posterior values shown in Figure \ref{cornerplot}, and integrate these stars forward for a time $t_{\text{opt}} = 83$ Myr. Figure \ref{posteriorcheck} compares these forward integrated model stars with the observed OCSN-49 stars in 6D phase space. Despite our relatively simple assumptions of sphericity and uncorrelated random birth velocities, the model shows good agreement in all dimensions, as it reproduces the bulk distribution of OCSN-49 observables. 

We note that our model predicts the existence of an extended tail towards smaller right ascensions. We expect that the \cite{HuntCatalog} catalog is incomplete, as the low-density features are a challenge for any algorithm, including \texttt{HDBSCAN}, to identify. We leave a dedicated search for additional extant cluster members for future study.

\section{Age Discrepancy} \label{age discrepancy}
 
We turn our attention to what makes OCSN-49 a particularly interesting case for studying the dynamical evolution of stars through the Milky Way: the discrepancy between its dynamical age ($\simeq$80 Myr) and its stellar age ($\simeq$600 Myr, though this age is more uncertain). Our dynamical model, which is able to explore ages up to (and greater than) 600 Myr finds no solution in this range, while the multiple methods described in Section~\ref{stellar ages} all indicate the stellar population of OCSN-49 is inconsistent with the dynamically derived age of 80 Myr. How then to resolve this apparent inconsistency?

The dynamical age of OCSN-49 is derived to be the time when the orbits of its stars converged, not necessarily the ages of the stars themselves. The two ages are naturally explained by the following scenario: OCSN-49 formed as a bound star cluster $\simeq$600 Myr ago. After traversing the galaxy for roughly 500 Myr it underwent some sort of destructive interaction that unbound its constituent stars. That disruptive encounter, combined with Galactic tidal forces, formed the stellar stream we see today roughly 80 Myr later. Under this scenario, OCSN-49 exists in a transition state, near the end of a star cluster's life as it disperses into the Milky Way field.

It is widely accepted that star clusters, especially those that reside in the disks of galaxies, do not live forever. A paucity of open clusters older than 0.5 Gyr was first noted by \cite{Oort_1958}, and the notion that open clusters are eventually disrupted has since been robustly confirmed observationally \citep[e.g.,][]{Wielen_1971, Lamers_2005, Tang_2019, Castro-Ginard_2020, Almeida_2025, Moreira_2025}, through theoretical arguments \citep[e.g.,][]{Spitzer_1958, Wielen_1985, Chernoff_1990, Lamers_2006}, and using $N$-body simulations of disruptive encounters \citep[e.g.,][]{Baumgardt_2003, Trenti_2010}. Along with stellar evolution, a variety of mechanisms may cause open clusters to lose mass and disperse their stars into the field, including tidal interactions with their host galaxy \citep[e.g.,][]{Bergond_2001, Dalessandro_2015}, spiral arm passages \citep{Gieles_2007}, and interactions with GMCs \citep{Gieles_2006}. For clusters in the solar neighborhood, disruptions by GMCs are expected to be the dominant disruptor \citep[see Figure 1 in][]{Lamers_2006}. 

In order to explore the possibility that OCSN-49 is the product of a disruptive encounter, we first derive a mass estimate for the cluster (Section \ref{mass}). Assuming that a single interaction between a GMC and OCSN-49 was responsible for its disruption, we derive a first estimate of the GMC mass and impact parameter in Section~\ref{disruption}. However, as we discuss below, multiple cluster-GMC encounters may have been responsible for donating energy to, and eventually unbinding, OCSN-49.

\subsection{Mass Estimate} \label{mass}
To obtain a mass estimate for OCSN-49, we fit the stream to a Kroupa initial mass function (IMF) \citep{KroupaIMF}. Gaia is essentially complete for stars with $ G < 17$ \citep{Gaia17mag}, so we solve for the normalization constant $N_{0}$:
\begin{equation}
n_{\text{obs}} = N_0 \int_{M_{\text{min}}}^{\infty} m^{\alpha} \, dm
\end{equation}
where $n_{\text{obs}}$ is the number of OCSN-49 stars observed with $ G < 17$, $M_{\text{min}}$ is the minimum mass, and $\alpha$ is the Kroupa IMF power law index ($\alpha$ = -0.3 for 0.01 M$_{\odot}$ $<  m < $ 0.08 M$_{\odot}$, 
$\alpha$ = -1.3 for 0.08 M$_{\odot}$ $<  m < $ 0.5 M$_{\odot}$,
$\alpha$ = -2.3 for $m > $ 0.5 M$_{\odot}$). We determine $M_{\text{min}}$ = 0.51 M$_{\odot}$ by obtaining the smallest FLAME mass for stars with $ G < 17$ in OCSN-49 using the FLAME (Final Luminosity and Age Estimator) module of the the astrophysical parameters inference system (Apsis) chain \citep{Flame1, Flame2}. 

To estimate the total number of stars, $n_{\text{total}}$ and total mass, $m_{\text{total}}$, we assume continuity of the IMF at crossover points when solving for the normalization constants, $N_{0}$ in each region. This procedure results in $n_{\text{total}}$ $\approx$ 530 and $m_{\text{total}}$ $\approx$ 225 M$_{\odot}$. Our inferred mass closely agrees with the mass estimate of 218.82 M$_{\odot}$ reported by \cite{HuntCatalog}, who fit the stream to a Kroupa IMF as well. We note that the initial stellar mass of OCSN-49 is likely somewhat larger as: 1) stellar evolution naturally causes clusters to lose mass, 2) stars were likely ejected dynamically from the cluster at birth and these would have long since migrated to other regions of the Galaxy, and 3) the \citet{HuntCatalog} catalog is likely incomplete, even within the magnitude limits of Gaia (for instance, the \texttt{HDBSCAN} algorithm used could be missing the tails in phase space predicted by our model). Nevertheless, we expect that our derived mass of 225 M$_{\odot}$ provides a reasonable estimate for our subsequent calculations.

\subsection{The Disruption of OCSN-49: A Single Disruption via Giant Molecular Cloud?}
\label{disruption}

The discrepancy between the dynamical age derived in Section \ref{model} ($\simeq$80 Myr) and the stellar age ($\simeq$600 Myr) suggests that a disruption event occurred roughly $520$ Myr into OCSN-49's lifetime. This timescale provides a powerful comparison to theoretical models of cluster disruption. As an estimate for the disruption timescale via a single cluster-GMC interaction for OCSN-49, we utilize the analytic formulation derived in \cite{Gieles_2006}:
\begin{equation}\label{eq: t_single}
    t_{\rm dis}^{\rm single} = \kappa \rho_{\rm n}^{-1} \sqrt{M_{\rm c} / r_{\rm h}^3}
\end{equation}
where $\kappa$ is a numerical factor defined in \cite{Gieles_2006}, $\rho_{\rm n}$ is the global density of GMCs, $M_{\rm c}$ is the cluster mass, and $r_{\rm h}$ is the half-mass radius of the cluster. We utilize the result of \cite{Larsen_2004}, which relates the half-mass radius to the cluster mass, to estimate the pre-interaction half-mass radius ($r_{\rm h} \approx 3$ pc, assuming $M_{\rm c} = 225 ~\rm M_{\odot}$ as derived in Section \ref{mass}). We use $\rho_{\rm n} = 0.03$ M$_{\odot}$ pc$^{-3}$ \citep{Solomon_1987} as the global density of GMCs in the solar neighborhood. Under these assumptions, we find $t_{\rm dis}^{\rm single} = 265$ Myr. While this timescale is a factor of two smaller than the observed $\simeq$500 Myr age discrepancy, we emphasize that this is an order-of-magnitude estimate, as the timescale in Equation \ref{eq: t_single} is valid as a statistical average; individual clusters will have differing lifetimes due to the stochastic nature of GMC encounters \citep{Gieles_2006}. 

We can compare this time of 265 Myr to estimates in the literature for the timescale of interaction with a GMC. For instance, \cite{Kokaia_2019} find that the Sun should pass through a GMC every $\sim625$ Myr. Since these authors assume a direct collision, it serves as an approximate upper bound for the disruption of a low-mass cluster on a Sun-like orbit such as OCSN-49 which can disrupt even when a GMC passes some distance away. Additionally, \cite{Moreira_2025} infer the open cluster-GMC destructive encounter rate by reproducing the observed scale height of nearby OCs, determining that an open cluster on a solar orbit should be destroyed by a GMC in $\simeq 700$ Myr. That our observed timescale of roughly $500$ Myr sits beneath these estimates is encouraging.

We can go a step further, deriving the likely characteristics of the GMC that disrupted OCSN-49.
\cite{Gieles_2006} derive an approximate expression for the energy injection via a single cluster-GMC interaction:
\begin{equation}\label{eq: delta_E}
    \Delta E \simeq \frac{4.4 r_{\rm h}^2}{(p^2 + \sqrt{r_{\rm h} R_{\rm n}^3})^2} {\bigg( \frac{G M_{\rm n}}{V_{\rm max}} \bigg)}^2 M_{\rm c}
\end{equation}
where $p$ is a slight variation on the typical impact parameter, defined to be the closest approach distance between the cluster and GMC, $R_{\rm n}$ is the radius of the GMC, $M_{\rm n}$ is the GMC mass, and $V_{\rm max}$ is the maximum relative velocity between the cluster and GMC. The parameters $p$ and $V_{\rm max}$ are analogous to the traditional impact parameter $b$ and relative velocity at infinity $V_{\infty}$, but are modified to account for gravitational focusing. It is well established that a mass-size relation in GMCs exists \citep[e.g.,][]{Larson_1981, Lombardi_2010, Lada_2020} due to the log-normal column density that GMCs exhibit. We use the scaling relation from \cite{Chen_2020}, i.e. $M_{\rm n} = 156.6 (R_{\rm n} / \rm{pc})^{1.96} ~\rm M_{\odot}$.

There are then three parameters independent of the cluster properties that describe the encounter: the distance of closest approach $p$, the impact velocity $V_{\rm max}$, and the mass of the GMC, $M_{\rm n}$. \cite{Gieles_2006} show, through their relation of energy injection to mass loss, that destructive encounters in the solar neighborhood occur roughly when the injected energy is equivalent to four times the total cluster energy. We estimate the total cluster energy
as $E_{0, \rm c} = -\eta G M_{\rm c}^2 / 2 r_{\rm h} = -3.8 \times 10^{44} ~\rm ergs$, where $\eta \approx 0.4$ for a Plummer model.

\begin{figure}
\begin{center}
   \includegraphics[width=1.0\columnwidth]{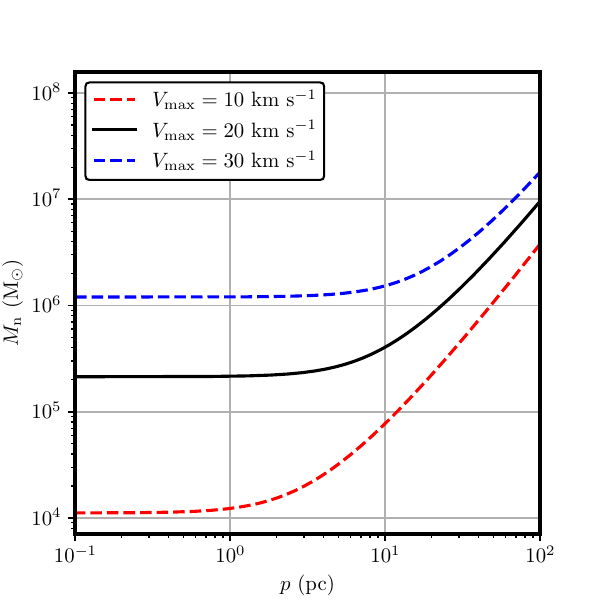}
   \caption{The minimum GMC mass $M_{\rm n}$ versus impact parameter $p$ required to produce a destructive encounter with $\Delta E / |E_{0, \rm c}| \geq 4$, plotted at three different relative velocities ($V_{\rm max}$). The solid black line, with $V_{\rm max} = 20 $ km s$^{-1}$, is our fiducial value. The impact parameters $p$ and $V_{\rm max}$ are analogous to the traditional impact parameters $b$ and $V_{\infty}$. }\label{pvsmass}
\end{center}
\end{figure}

In Figure \ref{pvsmass}, we show for three constant values of $V_{\rm max}$ the minimum GMC mass $M_{\rm n}$ as a function of $p$ to generate a destructive encounter with $\Delta E / |E_{0, \rm c}| \geq 4$. Every encounter in $M_{\rm n}-p$ space above each respective line would be destructive to OCSN-49 at that given $V_{\rm max}$. The behavior in Figure \ref{pvsmass} is expected; lower relative velocities between the cluster and GMC lead to less massive GMCs being able to destroy the structure, while larger impact parameters require more massive GMCs for a disruptive encounter. 

To find the parameters of the most likely destructive single encounter, we must consider the distributions of the encounter parameters $M_{\rm n}$, $p$, and $V_{\rm max}$. We make the simplifying assumption that $V_{\rm max} = 20$ km s$^{-1}$, chosen to be twice the relative velocity dispersion between clusters and GMCs in the solar neighborhood \citep[for an in-depth explanation, see Section 5.1 in][]{ Gieles_2006, Stark_1984, Piskunov_2006}. Under the assumption that GMCs are homogeneously distributed throughout the disk, we maximize the joint probability distribution $P(M_{\rm n}, p) = P(M_{\rm n})P(p)$ under the constraint that $\Delta E \geq 4 |E_{0, \rm c}| $.  For our GMC mass and impact parameter distributions, we use:
\begin{equation}\label{eq: priors}
  \begin{aligned}
    P(M_{\rm n}) & = M_{\rm n}^{-2} \\
    P(p)  & = p \\
  \end{aligned}
\end{equation}
where we ignore any normalizing constants. Our GMC mass function comes both from theory \citep[e.g.,][]{Fleck_1996, Wada_2020} and observation \citep[e.g.,][]{Rice_2016, Mok_2020}. Our impact parameter distribution comes from the simple assumption that the differential cross-sectional area for a potential interaction scales as $dA = 2 \pi p dp$. 

\begin{figure}
\begin{center}
   \includegraphics[width=1.0\columnwidth]{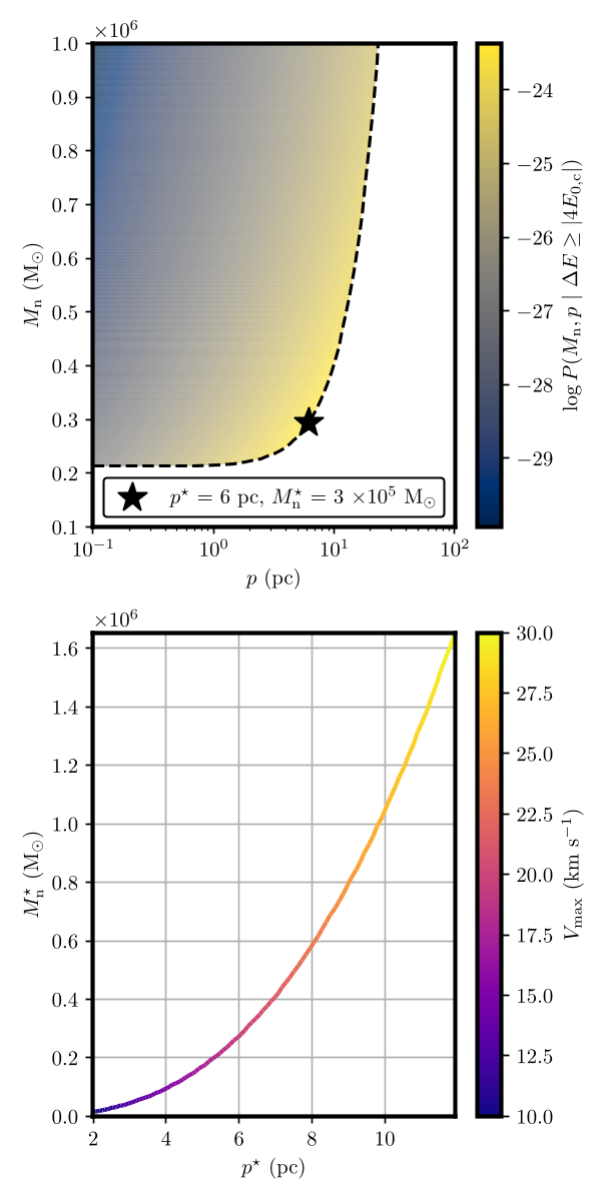}
   \caption{\textit{Top}: GMC mass $M_{\rm n}$ $-$ encounter parameter $p$ space, colored by the un-normalized log-likelihood of the joint probability distribution $P(M_{\rm n}, p) = P(M_{\rm n})P(p)$, assuming $V_{\rm max} = 20$ km s$^{-1}$. The dashed black line represents $\Delta E = 4 |E_{0, \rm c}|$. Our most likely perturber solution is labeled as a star. \textit{Bottom}: The most likely GMC mass $M_{\rm n}^{\star}$ vs the most likely encounter parameter $p^{\star}$ for a given maximum relative velocity $V_{\rm max}$, colored by $V_{\rm max}$. Encounters at low relative velocities prefer very close interactions with a less massive GMC, while the opposite is true at higher relative velocities. }\label{perturber_M_p}
\end{center}
\end{figure}

Following these assumptions, we find the most likely parameters for the GMC that disrupted OCSN-49:
$$M_{\rm n}^{\star} = 3 \times 10^5 ~ \rm{M_{\odot}} $$
$$p^{\star}  = 6 ~ \rm{pc} $$
Had OCSN-49 been destroyed in a single encounter, it would have likely been due to a nearly head-on collision with a fairly massive GMC. In Figure \ref{perturber_M_p}, we show the relative log-likelihoods across a grid of encounter scenarios in $M_{\rm n}-p$ space, and over-plot our model's preferred solution in the top panel. More probable solutions lie very close to the line $\Delta E = 4 |E_{0, \rm c}|$. In the bottom panel of Figure \ref{perturber_M_p}, we allow $V_{\rm max}$ to vary and solve for the most likely GMC mass $M_{\rm n}^{\star}$ and distance of closest approach $p^{\star}$ at that given $V_{\rm max}$. We note that lower relative velocities between the GMC and OCSN-49 prefer very close encounters with lower mass GMCs, while the opposite is true at higher relative velocities. 

It should be emphasized that OCSN-49 may have undergone successive GMC encounters that led eventually to its unbinding. Indeed, studies of open cluster disruption \citep{Spitzer_1958, Gieles_2006} suggest that multiple encounters are statistically likely to have played a role. Quantifying these effects is outside the scope of this work, as a complete analysis would self-consistently evolve the cluster's internal structure through these encounters within a Milky Way potential. Furthermore, the cluster may also have been affected by other disruption processes, such as tidal shocks and interactions with the Milky Way's spiral arms. Nevertheless, it is heartening that even with our approximate model, we find reasonable parameters for the derived GMC interaction and a disruption time that aligns well with theoretical expectations.

\section{Conclusion} \label{conclusion}
We have carefully analyzed the membership, chemistry, kinematics, and dynamical origin of OCSN-49. We produce a slightly revised catalog of 257 stars, with Gaia radial velocities for 93 stars. We present and utilize Gemini-N/MAROON-X high-resolution spectra for four stars across the extent of the structure, and derive up to 15 precise elemental abundances for each star. These stars are consistent with solar metallicity, and show remarkable homogeneity across the derived abundances, further supporting OCSN-49's coeval nature. 

We derive an age for the cluster using three independent methodologies. Using \texttt{Brutus}, a Python package designed to simultaneously determine age, metallicity, reddening, extinction, and distance to a stellar population via isochrone fitting, we find an age of $\simeq$600 Myr, which we show to be robust across multiple regions of the structure. With our derived lithium abundances, we make comparisons to well-studied open clusters across a range of ages, and find that the lithium content in OCSN-49 is most consistent with the 420 Myr old M48. However, additional data are required to derive more stringent conclusions. We measured rotation periods for 110 members with TESS and ZTF light curves and likewise found that its gyrochronological age is inconsistent with the Pleiades and is closer to 500--600~Myr. We backtrack orbits of stars with sufficiently precise Gaia radial velocities, and find that the ends of the structure become compact in position space roughly 130 Myr ago. We utilize a slightly modified version of the forward-modeled dynamical age-dating method presented by \cite{Kyle}, and determine an age of $83 \pm 1$ Myr. Via posterior predictive checking, we see the model is able to accurately reproduce the bulk characteristics of OCSN-49, agreeing in all phases of 6D parameter space. 

These independent age indicators point to a clear discrepancy between the models reliant on stellar evolution, which support an older age, and our dynamical age-dating technique, which supports a younger age. This discrepancy is naturally explained if OCSN-49 is indeed an older population of stars that traversed the Milky Way as a bound open cluster before undergoing a destructive interaction roughly 80 Myr ago, unbinding the structure and eventually forming the disperse stellar stream we see today. Using star counts and considering \gaia's sensitivity, we derive a stellar mass of $\approx$ 225 M$_{\odot}$ for OCSN-49. Motivated by our disruption hypothesis, we explore a scenario in which OCSN-49 was disrupted by a single interaction with a GMC to place analytic constraints on such an encounter. We find that this scenario predicts a cluster disruption timescale of $\simeq$250 Myr, a factor of two faster than the $\simeq$500 Myr timescale that our age discrepancy suggests, but consistent at an order-of-magnitude level. The observed age discrepancy also aligns well with rate predictions for open cluster-GMC interactions for an open cluster on a Sun-like orbit \citep[e.g.][]{Kokaia_2019, Moreira_2025}. We estimate the energy needed to unbind OCSN-49, and marginalize over the distributions of GMC masses and impact parameters, finding that a nearly head-on collision with a fairly massive GMC ($\sim$ $10^5$ M$_{\odot}$) is the most likely scenario, under the assumption that OCSN-49 was disrupted by a single GMC encounter. 

Given the availability of 6D phase space information for many of the stars in OCSN-49, this structure is ripe for a more detailed follow-up study. $N$-body modeling of a cluster-GMC encounter, aimed at best reproducing the distribution of positions and velocities observed today, could place tight constraints on not only the encounter parameters ($M_{\rm n}$, $p$, $V_{\rm max}$) described above, but could also indicate the relative direction and angle at which the encounter occurred. We leave such a study for future work.

Our results demonstrate that the combination of model-dependent age-dating techniques with dynamical modeling is a valuable tool that allows us to explore the theory of cluster disruption in the \gaia era. Objects like Meingast-1, Theia 456, and OCSN-49 are just the first to be considered under this lens. As our astrometric data improves, more of these disperse populations will inevitably be discovered, helping us bridge the gap between our current understanding of star formation and the population of the Milky Way field.

\section*{Acknowledgment}

We thank the anonymous referee for their careful reading of the manuscript and comments that improved its quality.

This manuscript is the result of a Research Experiences for Undergraduates program at the University of Florida Physics Department, supported by the National Science Foundation (NSF), United States via DMR-2244024. A.N.M. gratefully acknowledges support from this program.  S.C.S. was supported by a Research Innovation and Scholarly Excellence (RISE) Grant from the University of Tampa. This work was also supported by NASA TESS GI grant No.\ 80NSSC24K0501 and 80NSSC24K0502 (program IDs G06163 and G06164). J.C. acknowledges support from the Agencia Nacional de Investigación y Desarrollo (ANID) via Proyecto Fondecyt Regular 1191366 and 1231345, and by ANID BASAL project FB210003.

This work is based on observations obtained at the international Gemini Observatory, a program of NSFs NOIRLab, which is managed by the Association of Universities for Research in Astronomy (AURA) under a cooperative agreement with the National Science Foundation on behalf of the Gemini Observatory partnership: the National Science Foundation (United States), National Research Council (Canada), Agencia Nacional de Investigación y Desarrollo (Chile), Ministerio de Ciencia, Tecnología e Innovación (Argentina), Ministério da Ciência, Tecnologia, Inovações e Comunicações (Brazil), and Korea Astronomy and Space Science Institute (Republic of Korea). These observations were obtained under program GN-2023B-FT-201.

We thank Andreas Seifahrt and the other members of the MAROON-X support team at the University of Chicago for their careful reduction of the spectra used in this study. We also thank Constantine Deliyannis for providing the $\lambda 6707$ region linelist.

\facilities{Gemini:Gillett (MAROON-X), Gaia, TESS, PO:1.2m (ZTF)}

\software{\texttt{AstroPy} \citep{Astropy}, 
\texttt{astroquery} \citep{astroquery}, \texttt{Brutus}\footnote{See footnote 1.}, \texttt{Corner} \citep{corner}, \texttt{emcee} \citep{emcee}, \texttt{gala} \citep{Gala}, \texttt{galpy} \citep{galpy}, \texttt{Matplotlib} \citep{Matplotlib}, \texttt{MOOG} \citep{Moog} \texttt{NumPy} \citep{Numpy},
\texttt{ptemcee} \citep{ptemcee}, \texttt{SciPy} \citep{Scipy}, \texttt{Seaborn} \citep{Seaborn}, \texttt{SPAE} \citep{Schuler}, \texttt{Pandas} \citep{pandas}, \texttt{TESScut} \citep{TESScut}, \texttt{unpopular} \citep{Hattori2022}, the IDL Astronomy User's Library \citep{IDLastro}}

\bibliography{main}{}
\bibliographystyle{aasjournal}

\end{document}